\documentclass[aps,prd,a4paper,twocolumn,amsmath,showpacs,superscriptaddress,nofootinbib,preprintnumbers]{revtex4-1}

\usepackage{verbatim}
\usepackage{epsfig}
\usepackage{graphicx}
\usepackage{multirow}
\usepackage{dcolumn}
\usepackage{amsmath}
\usepackage{mathtools}
\usepackage{amsfonts}
\usepackage{amssymb}
\usepackage{epstopdf}
\usepackage{bm}
\usepackage{siunitx}
\usepackage{braket}
\usepackage{enumitem}
\usepackage{soul}
\usepackage{xcolor}

\definecolor{navyblue}{rgb}{0.0, 0.0, 0.5}
\definecolor{royalblue}{rgb}{0.25, 0.41, 0.88}
\definecolor{cadmiumgreen}{rgb}{0.0, 0.42, 0.24}
\definecolor{blue-violet}{rgb}{0.54, 0.17, 0.89}
\definecolor{darkviolet}{rgb}{0.58, 0.0, 0.83}
\definecolor{orange(colorwheel)}{rgb}{1.0, 0.5, 0.0}

\usepackage{hyperref}
\hypersetup{
    colorlinks=true, 
    linkcolor=royalblue, 
    citecolor=magenta}

\newcommand\ee{\end{equation}}
\newcommand\be{\begin{equation}}
\newcommand\eea{\end{eqnarray}}
\newcommand\bea{\begin{eqnarray}}
\newcommand\mpl{M_{\rm P}}

\newcommand\gev{\mathrm{GeV}}
\newcommand\mev{\mathrm{MeV}}

\newcommand\mpc{\mathrm{Mpc}}

\newcommand\hz{\mathrm{Hz}}
\newcommand\K{\mathrm{K}}

\newcommand\planck{\emph{Planck}}
\newcommand\planckBKP{\emph{Planck} + BKP}
\newcommand\planckBKPEXT{\emph{Planck} + BKP + EXT}
\newcommand\planckBKPFIRAS{\emph{Planck} + BKP + FIRAS}
\newcommand\planckBKPpulsar{\emph{Planck} + BKP + pulsar}
\newcommand\planckBKPLIGO{\emph{Planck} + BKP + LIGO-Virgo}

\renewcommand{\mpl}{M_{\rm P}}
\newcommand{\nt}{n_{\rm t}}

\newcommand{\ns}{n_{\rm s}}

\newcommand{\neff}{N_{\rm eff}}
\newcommand{\ngw}{N^{\rm GW}_{\rm eff}}
\newcommand{\nbbn}{N^{\rm GW}_{\rm eff,BBN}}
\newcommand{\ncmb}{N^{\rm GW}_{\rm eff,CMB}}
\newcommand{\fsky}{f_{\rm sky}}

\newcommand\limit[1]{#1\%\,\mathrm{CL}}
\newcommand{\lcdm}{\Lambda\mathrm{CDM}}

\newcommand{\tblack}[1]{\textcolor{black}{#1}} 

\newcommand{\tu}{\textup}
\newcommand{\dif}{\mathrm{d}}
\newcommand\ie{{\it i.e.}~}
\newcommand\eg{{\it e.g.}~}

\newcommand\eq[1]{Eq.~\eqref{eq:#1}}
\newcommand\sect[1]{Sec.~\ref{sec:#1}}
\newcommand\fig[1]{Fig.~\ref{fig:#1}}
\newcommand\tab[1]{Tab.~\ref{tab:#1}}
\newcommand{\Tr}{\mathrm{Tr}}

\newcommand\vertsp{\rule[-2mm]{1mm}{0mm} &}
\newcommand\horsp{\rule[-2mm]{0mm}{5.5mm}}

\begin{document}

\title{Updated Constraints and Forecasts on 
Primordial Tensor Modes}

\author{Giovanni Cabass}
\affiliation{Physics Department and INFN, Universit\`a di Roma 
	``La Sapienza'', P.le\ Aldo Moro 2, 00185, Rome, Italy}

\author{Luca Pagano}
\affiliation{Physics Department and INFN, Universit\`a di Roma 
	``La Sapienza'', P.le\ Aldo Moro 2, 00185, Rome, Italy}

\author{Laura Salvati}
\affiliation{Physics Department and INFN, Universit\`a di Roma 
	``La Sapienza'', P.le\ Aldo Moro 2, 00185, Rome, Italy}

\author{Martina Gerbino}
\affiliation{The Oskar Klein Centre for Cosmoparticle Physics, Department of Physics, Stockholm University, AlbaNova, SE-106 91 Stockholm, Sweden}
\affiliation{Nordita (Nordic Institute for Theoretical Physics), Roslagstullsbacken 23, SE-106 91 Stockholm, Sweden}
\affiliation{Physics Department and INFN, Universit\`a di Roma 
	``La Sapienza'', P.le\ Aldo Moro 2, 00185, Rome, Italy}

\author{Elena Giusarma}
\affiliation{McWilliams Center for Cosmology, Department of Physics, Carnegie Mellon University, Pittsburgh, PA 15213, USA.}
\affiliation{Physics Department and INFN, Universit\`a di Roma 
	``La Sapienza'', P.le\ Aldo Moro 2, 00185, Rome, Italy}
	
\author{Alessandro Melchiorri}
\affiliation{Physics Department and INFN, Universit\`a di Roma 
	``La Sapienza'', P.le\ Aldo Moro 2, 00185, Rome, Italy}

\begin{abstract}
\noindent We present new, tight, constraints 
on the cosmological background of gravitational waves (GWs) using the latest 
 measurements of CMB temperature and polarization anisotropies provided by the \emph{Planck},  BICEP2 and \emph{Keck Array} experiments.  These constraints are further improved when the GW contribution $\ngw$ to the effective number of relativistic degrees of freedom $\neff$ is also considered. Parametrizing the tensor spectrum as a power law with tensor-to-scalar ratio $r$, tilt $\nt$ and pivot $0.01\,\mpc^{-1}$, and assuming a minimum value of $r=0.001$, we find $r < 0.089$, $\nt = 1.7^{+2.1}_{-2.0}$ ($\limit{95}$, no $\ngw$) and $r < 0.082$, $\nt = -0.05^{+0.58}_{-0.87}$ ($\limit{95}$, with $\ngw$). 
When the recently released $95\,\mathrm{GHz}$ data from \emph{Keck Array} are added to the analysis, the constraints on $r$ are improved to $r < 0.067$ ($\limit{95}$, no $\ngw$), $r < 0.061$ ($\limit{95}$, with $\ngw$). 
We discuss the limits coming from direct detection experiments such as LIGO-Virgo, pulsar timing (European Pulsar Timing Array) and  CMB spectral distortions (FIRAS). 
Finally, we show future constraints achievable from  a COrE-like mission:  if the tensor-to-scalar ratio is of order $\num{d-2}$ 
and the inflationary consistency relation $\nt = -r/8$ holds, COrE will be able to constrain $\nt$ with an error of $0.16$  at $\limit{95}$. In the case that lensing $B$-modes can be subtracted to $10\%$ of their power, a feasible goal for 
COrE, these limits will be improved to $0.11$ at $\limit{95}$. 
\end{abstract}


\maketitle

\twocolumngrid

\section{Introduction}
\label{sec:introduction}

\noindent After the \textcolor{black}{impressive} confirmation of the standard 
$\lcdm$ model of structure formation by many ground, balloon and space experiments \cite{deBernardis:2000gy, Eisenstein:2005su, Hinshaw:2012aka, Adam:2015rua}, the search for primordial gravitational waves (GWs) is one of the main goals of modern cosmology. Long-wavelength gravitational waves are predicted by the current go-to theory for the solution to the 
horizon and flatness problems of the 
Hot Big Bang picture (and the generation of primordial density perturbations), \ie cosmic inflation \cite{Guth:1980zm, Linde:1981mu, Albrecht:1982wi}. The scale at which inflation occurs is related to the amount of primordial tensor power, while the simplest models (with a single degree of freedom 
parametrizing the evolution of the inflationary energy density) 
predict 
the tensor spectrum to be slightly red-tilted (see \cite{Baumann:2009ds} for a comprehensive review). Constraining the tensor amplitude and tilt, 
then, will be an important step in the discrimination between different models of the 
early universe (see e.g. \cite{kamion} for a recent review).

In this paper we 
update the constraints on the parameters describing the tensor spectrum $P_\tu{t}(k)$ in light of the \emph{Planck} 2015 data \cite{Aghanim:2015xee}, 
the measurement of the $BB$ spectrum from the 
BICEP2/\emph{Keck}-\emph{Planck} (hereafter BKP) 
combined analysis \cite{Ade:2015tva}, \tblack{and the recently released $95\,\mathrm{GHz}$ data from \emph{Keck Array} \cite{Array:2015xqh}}. We consider a power law parametrization of $\Delta^2_\tu{t}(k)\equiv k^3P_\tu{t}(k)/2\pi^2$ in terms of the 
tensor amplitude $A_\tu{t}$ and the tilt $\nt$, 
\ie 
\begin{equation}
\label{eq:deltat}
\Delta^2_\tu{t}(k) = 
A_\tu{t}\bigg(\frac{k}{k_\star}\bigg)^{\nt}\,\,,
\end{equation}
where $A_\tu{t}$ is the amplitude of the 
tensor power spectrum 
at the pivot scale $k_\star$. The scalar spectrum $\Delta^2_\tu{s}(k)$ will be analogously parametrized in terms of its amplitude and the spectral index $\ns$, \ie 
\begin{equation}
\label{eq:deltas}
\Delta^2_\tu{s}(k) = A_\tu{s}\bigg(\frac{k}{k_\star}\bigg)^{\ns - 1}\,\,,
\end{equation}
and we will consider the tensor-to-scalar ratio $r\equiv A_\tu{t}/A_\tu{s}$ in place of $A_\tu{t}$ in the following analysis. 

It is well known that the sensitivity of 
Cosmic Microwave Background (CMB) experiments 
to $P_\tu{t}(k)$ comes from the contribution of primordial tensor modes to the angular power spectra of the 
CMB temperature and polarization anisotropies, \ie
\begin{equation}
\label{eq:Cl-tens}
C^{XY, \text{t}}_\ell = \int_0^{+\infty}\frac{\dif k}{k} \Delta^\tu{t}_{\ell,X}(k) \Delta^\tu{t}_{\ell,Y}(k) P_\tu{t}(k)\,\,,
\end{equation}
where $\Delta^\tu{t}_{\ell,X}(k)$ are the transfer functions ($X = T, E, B$) for tensor modes, dependent on late-time physics. The accurate measurement of $T$, $E$ and $B$ anisotropies from the \emph{Planck} experiment at large scales, together with the limit on $B$-mode polarization from the BKP joint analysis, has allowed to obtain $r_{0.002} < 0.08$ at $\limit{95}$ 
(\emph{Planck} TT + lowP + BKP dataset: see \cite{Ade:2015lrj}).

Measurements of CMB anisotropies alone, however, are limited 
to scales $k$ from 
$\approx 10^{-3}\,\mpc^{-1}$ to 
$\approx 10^{-1}\mpc^{-1}$: a much larger range of scales 
becomes available when one considers 
the following 
experiments, also sensitive to 
primordial GWs \cite{Allen:1996vm, Maggiore:1999vm, Cutler:2002me} (see figure A2 of \cite{Moore:2014lga} for details about sensitivities) :
\begin{itemize}[leftmargin=*]
\item direct detection experiments, such as LIGO \cite{Abbott:2007kv, Abadie:2011xta}, 
and Virgo \cite{Acernese:2004jn, TheVirgo:2014hva}. 
With these ground-based interferometers 
one can probe the primordial gravitational wave spectrum in a range of frequencies $\Delta f$ (with $2\pi f = c k$) from $\sim 1\,\hz$ to $\sim 10^4\,\hz$, while the planned space-based 
eLISA \cite{AmaroSeoane:2012km, AmaroSeoane:2012je}, DECIGO \cite{Kawamura:2006up, Kawamura:2011zz} and the proposed Big-Bang Observer (BBO) \cite{Crowder:2005nr} focus on frequencies from $\sim 10^{-4}\,\hz$ to $\sim 1\,\hz$; 
\item high-stability pulsar timing experiments, 
like the European Pulsar Timing Array (EPTA) \cite{Ferdman:2010xq}, which are sensitive to GWs in frequencies between $\sim 10^{-9}\,\hz$ and $\sim 10^{-7}\,\hz$;
\item measurements of the CMB energy spectrum 
\cite{Fixsen:1996nj}: two recent papers \cite{Ota:2014hha, Chluba:2014qia} showed how the integrated tensor power from $k\approx 10^3\,\mpc^{-1}$ ($f\approx 10^{-12}\,\hz$) to $k\approx 10^6\,\mpc^{-1}$ ($f\approx 10^{-9}\,\hz$) gives a contribution to the spectral distortions of the CMB spectrum, subleading with respect to distortions caused by Silk damping of acoustic waves in the baryon-photon fluid \cite{Hu:1992dc, Chluba:2012gq, Pajer:2013oca} (which, conversely, allow to probe the integrated spectrum of scalar perturbations \cite{Hu:1994bz, Chluba:2012we}).
\end{itemize}

\begin{figure}[!hbtp]
\includegraphics[width=0.5\textwidth]{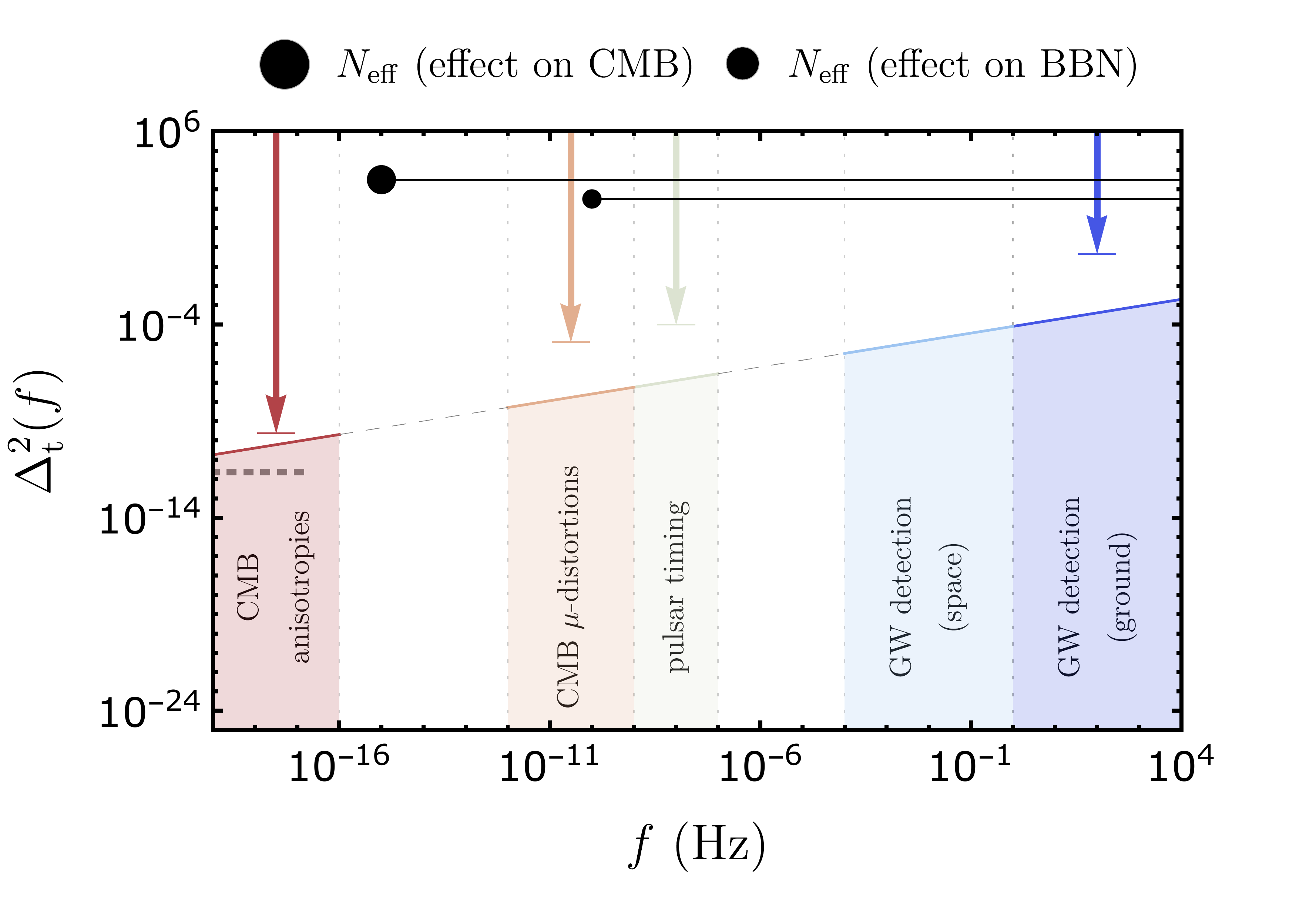}
\caption{\footnotesize{Cartoon plot of a power-law (blue) primordial tensor spectrum ($\Delta^2_\tu{t} = \num{d-10}$, $\nt = 0.35$) over a range of frequencies $f$ going from $10^{-17}\,\hz$ (smallest frequency that can be probed with CMB anisotropies) to $10^4\,\hz$ (largest frequency available to ground-based experiments). The vertical arrows represent the current upper bounds on the tensor amplitude at different scales. We show with a gray dotted line the prior on $r$ used in our analysis ($r > 0.001$). The filled regions show 
which is the relevant $\Delta f$ for the various observables discussed in the text. We refer to \sect{observables-nucleosynthesis} for a more accurate discussion regarding 
primordial abundances and the effect on CMB anisotropies. We stress that this plot has only illustrative purposes.}}
\label{fig:scales}
\end{figure}

For a visual impression of the various scales probed by these different observables see the illustrative plot \fig{scales}.

In addition to this, primordial gravitational waves have also an effect on the expansion of the universe: being relativistic degrees of freedom, they will add to the effective number of relativistic species $\neff$ \cite{Mangano:2001iu, Mangano:2005cc}, increase the radiation energy density, and therefore decrease the redshift of matter-radiation equality, as one can see from the relation
\begin{equation}
\label{eq:zeq}
1 + z_\tu{eq} = \frac{\Omega_\tu{m}}{\Omega_\tu{r}} = \frac{\Omega_\tu{m}}{\Omega_\gamma(1 + 0.2271\neff)}\,\,. 
\end{equation}
This alters the CMB power spectrum in a way similar to that of additional sterile, massless neutrinos (see \cite{Bowen:2001in, Hou:2011ec} and references therein for an analysis of these effects).

The contribution to the radiation content of the universe will also affect primordial nucleosynthesis (BBN). A larger amount of gravitational waves will result in a more rapid expansion of the universe: this, in turn, means that neutrons will have less time to decay before the freeze-out of weak interactions, leading to a larger neutron-to-proton ratio 
and an overproduction of helium. 
One can then consider the astrophysical constraints on the abundances of light elements like helium \cite{Izotov:2014fga, Mucciarelli:2014abc} and Deuterium \cite{Cooke:2013cba}
, and the effect that altering the value of the primordial helium abundance $Y_P$ has on the CMB angular power spectrum \cite{Cyburt:2001pq, Ichikawa:2007js}.

Several authors have used these observables to provide constraints on both the primordial and not primordial (\eg the possible contribution coming from networks of cosmic strings \cite{Damour:2000wa, Olmez:2010bi}) gravitational wave spectrum \cite{Ungarelli:2005qb, Smith:2006xf, Chongchitnan:2006pe, Smith:2006nka, Boyle:2007zx, Stewart:2007fu, Camerini:2008mj,
DiValentino:2011zz, Sendra:2012wh, Gerbino:2014eqa, Henrot-Versille:2014jua, Meerburg:2015zua, Huang:2015gka, Nakama:2015nea, Pagano:2015hma, Huang:2015gca}. We note that in most of these works the tensor-to-scalar ratio $r$ is fixed at a reference value: if this value is high enough (\eg of order $\num{d-1}$), this will result in more stringent constraints on the tilt (see, \textit{e.g.}, \cite{Stewart:2007fu}). In our case, instead, we consider $r$ as a free parameter, varying along with the tilt. 

Before going on, we 
highlight which are the main novelties of this work: 
\begin{itemize}[leftmargin=*]
\item we include the recent $B$-mode polarization data coming from the BICEP2 and \emph{Keck Array} experiments \cite{Array:2015xqh} that improve significantly the constraints on the tensor mode amplitude $r$;
\item we examine which limits can be obtained from CMB $\mu$-distortions. While the current sensitivity (the state of the art being the FIRAS instrument on the COBE satellite) is too low for this observable to be competitive with the other ones discussed in the text, we note that this is a ``pure CMB'' constraint on the tilt, in the sense that no other observable besides the Cosmic Microwave Background is used to obtain it. Besides, one has to keep in mind that future experiments like PIXIE \cite{Kogut:2011xw} and LiteBIRD \cite{Matsumura:2013aja} are planned to have order \num{d3} times the sensitivity of FIRAS: therefore the limits from spectral distortions that we obtain in this paper will certainly be improved;
\item \textcolor{black}{we explicitly include in our analysis the above mentioned effect of gravitational waves on $Y_P$ 
and the bound on $\neff$ from the observations of light element abundances;} 
\item we perform a forecast 
on the parameters $r$ and $\nt$ 
combining future CMB experiments like COrE \cite{Bouchet:2011ck} and GW direct detection experiments as AdvLIGO \cite{TheLIGOScientific:2014jea}. 
We consider a fiducial cosmology 
where the tensor-to-scalar ratio is of order \num{d-2}, with tilt 
fixed by the single-field slow-roll consistency relation $\nt = -r/8$. 
Such values of $r$ will be probed by ground-based experiments like AdvACT 
\cite{Calabrese:2014gwa}, a new receiver for the Atacama Cosmology Telescope 
based on projected improvements of the existing ACTPol camera \cite{Niemack:2010wz}. 
The reason for this analysis is the following: if primordial tensor modes are detected, and there is no more the freedom of changing the overall scale of the spectrum, constraining its scale dependence will be one of the main goals of future $B$-mode cosmology \cite{footnote0}; 
\item we include delensing in our forecasts: recently it has been shown that lensing $B$-modes can be subtracted to $10\%$ of their original power if noise is brought down to $\approx 1\,\mu\K\cdot\mathrm{arcmin}$ \cite{Errard:2015cxa}. In this case, a larger range of multipoles becomes available to probe the scale dependence of the $B$-mode spectrum from tensors, leading to stronger constraints on $\nt$: we refer to \sect{method-delensing} 
for details.
\end{itemize}

The paper is organized as follows: 
the next section contains a more detailed description on the observables introduced above. In \sect{method} we present our data analysis and forecasting methods, in \sect{results} we discuss our results and we draw our conclusions in \sect{conclusions}.

\section{GW spectrum and observations}
\label{sec:observables}

\noindent The spectrum of primordial gravitational waves, at the present time and at a given frequency $f = k/2\pi$ (we take $a(\eta_0) = 1$), is given by \cite{Lidsey:1995np, Ungarelli:2005qb, Stewart:2007fu}
\begin{equation}
\label{eq:Omega-gwf}
\Omega_\tu{GW}(f)\equiv\frac{1}{\rho_c}\frac{\dif\rho_\tu{GW}}{\dif\log f} = \frac{\Delta^2_\tu{t}(f)}{24 z_\tu{eq}}\,\,,
\end{equation}
where $\rho_c = 3H^2/8\pi G$ is the critical density. This expression 
is found by solving the evolution equation for the gravitational wave amplitude in an expanding universe: for a detailed treatment of transfer functions for tensor perturbations, see \cite{Boyle:2005se, Watanabe:2006qe}.

Plugging in 
numerical values (\ie $f/\hz = \num{1.6d-15}k/\mpc^{-1}$), we obtain 
\begin{equation}
\label{eq:deltat-f}
\begin{split}
\Delta^2_\tu{t}(f) &= r A_\tu{s}\bigg(\frac{f}{f_\star}\bigg)^{\nt} \\
&= r A_\tu{s}\bigg(\frac{f/\hz}{\num{1.6d-17}}\bigg)^{\nt}\,\,.
\end{split}
\end{equation}
for a pivot scale $k_\star = 0.01\,\mpc^{-1}$.

Starting from this formula, we can discuss the impact that 
GWs have on the various observables that will be used in our analysis, starting from the effects on nucleosynthesis and the abundances of primordial elements.

\subsection{Nucleosynthesis and primordial abundances}
\label{sec:observables-nucleosynthesis}

\noindent The gravitational wave contribution to the number of relativistic degrees of freedom $g_*(T)$ at temperature $T$ is given by:
\begin{equation}
\label{eq:g-star}
g^{(\text{GW})}_*(T) = 2\bigg(\frac{T_\tu{GW}}{T_\gamma}\bigg)^4 = 2\frac{\rho_\tu{GW}}{\rho_\gamma}\,\,,
\end{equation}
where the factor of $2$ comes from the two helicities of tensor perturbations. At temperatures $T\gtrsim 1\,\mev$, when the relativistic degrees of freedom are 
$e^\pm$, $\gamma$, $\nu$ and GWs, expressing $g_*(T)$ in terms in the effective number of neutrino species $\neff = 3.046 + \ngw$ gives
\begin{equation}
\label{eq:Ngw-rho}
\ngw = \frac{8}{7}\frac{\rho_\tu{GW}}{\rho_\gamma}\,\,.
\end{equation}
In order to relate this to \eq{Omega-gwf}, which involves the spectrum of primordial gravitational waves at the present time, one notes that from $T\gtrsim1\,\mev$ to the present time $\rho_\tu{GW}$ scales as $a^{-4}$, while the photon energy density evolves as $\rho_\gamma\sim 1/(a^4 g_{*,s}^{4/3})$ (\ie by keeping entropy constant). The result is:
\begin{equation}
\label{eq:Ngw-rho-gstar}
\ngw = \frac{8}{7}\bigg(\frac{g_{*,s}(T\gtrsim1\,\mev)}{g_{*,s}(T_0)}\bigg)^{\frac{4}{3}}\frac{\rho_\tu{GW}}{\rho_\gamma}\bigg|_{\eta = \eta_0}\,\,,
\end{equation}
where $T_0$ is the temperature of the CMB at the present time $\eta = \eta_0$. Multiplying both $\rho_\tu{GW}$ and $\rho_\gamma$ by $1/\rho_c$, and using the definition of $\Omega_\tu{GW}(f)$ given in \eq{Omega-gwf}, one obtains:
\begin{equation}
\label{eq:Ngw-rho-gstar-f}
\ngw = \frac{8}{7}\bigg(\frac{g_{*,s}(T\gtrsim1\,\mev)}{g_{*,s}(T_0)}\bigg)^{\frac{4}{3}}\frac{\rho_c}{\rho_\gamma}\int^{+\infty}_0\dif\log f\,\Omega_\tu{GW}(f)\,\,.
\end{equation}
Inserting numerical values one can find \cite{Maggiore:1999vm, Smith:2006nka, Pagano:2015hma}:
\begin{equation}
\label{eq:Ngw-final}
\ngw\approx\frac{h^2_0}{\num{5.6d-6}}\int^{+\infty}_0\dif\log f\,\Omega_\tu{GW}(f),\,.
\end{equation}
If, instead of considering temperatures $T\gtrsim1\,\mev$, one starts from times around decoupling (relevant for how $\neff$ affects the CMB spectra), one can find that the expression of $\ngw$ in terms of $\rho_\tu{GW}$ is the same as that of Eqs.~\eqref{eq:Ngw-rho-gstar-f} and \eqref{eq:Ngw-final}.

In the above equations the gravitational wave spectrum is integrated over all frequencies, from $f = 0$ to $f = +\infty$. In reality there are both IR and UV cutoffs to this integral:
\begin{itemize}[leftmargin=*]
\item the IR cutoff comes from the fact that the only gravitational waves that contribute to the radiation energy density at a given time $\overline{\eta}$ are those inside the horizon at $\eta = \overline{\eta}$. In fact these are the ones that have begun to oscillate, and then propagate as massless modes \cite{Pritchard:2004qp, footnote1}. This means that we have to consider two different $\ngw$:
\begin{enumerate}[leftmargin=*]
\item the first one is $\nbbn$, the contribution of GWs to $\neff$ that will enter in the computation of the abundances of primordial elements. Its IR cutoff will be the frequency corresponding to the horizon size at the time of nucleosynthesis, \ie $\approx\num{d-12}\,\hz$. Actually we take the cutoff to be of order $\num{d-10}\,\hz$, since, realistically, the gravitational waves will need to oscillate for a while after entering the horizon before contributing to $\rho_\tu{rad}$, see for details figure 2 of \cite{Pritchard:2004qp} and \cite{Smith:2006nka,Smith:2008fea};
\item the second one is $\ncmb$, that will affect the CMB power spectra through its effect on the redshift of matter-radiation equality and the comoving sound horizon. Its IR cutoff will be the horizon size at decoupling. This would be given by 
$\approx\num{d-17}\,\hz$, but we take it to be of order $\num{d-15}\,\hz$ for the same reason as before \cite{Smith:2006nka}.
\end{enumerate}
The two contributions to $\neff$ will differ a lot from each other only in the case of very red tensor spectra (when the resulting $\ngw$ is too small to have an effect anyway): for blue spectra the dependence of $\rho_\tu{GW}$ on the IR cutoff is very weak;
\item the UV cutoff is more arbitrary: if gravitational waves are produced by inflation, we expect a cutoff corresponding to the horizon size $k_\tu{end}$ ($\approx\num{d23}\,\mpc^{-1}$ for GUT-scale inflation and instant reheating - see \sect{appendix-reheating} for a derivation) at the end of inflation, since GWs of smaller wavelength will not be produced. 
Other authors make different choices for $UV$ cutoff, without referring to the inflationary theory \cite{Meerburg:2015zua}: for example one can suppose to have a production of gravitational waves up 
to the horizon size before the $\approx 60$ e-folds 
of hot Big Bang expansion. 
In \cite{Stewart:2007fu}, instead, the authors choose the UV cutoff to be given by the Planck frequency, \ie 
$f_\tu{P}\approx1/t_\tu{P} = \num{d43}\,\hz$ (with 
$k_\tu{P}\approx\num{d57}\,\mpc^{-1}$). 
Among these options, we choose 
$k_\tu{UV} = k_\tu{end}\approx\num{d23}\,\mpc^{-1}$: 
this allows 
to be more conservative with the constraints on the tensor tilt, since for a given $\nt$, a larger UV cutoff will result in a larger $\ngw$. \textcolor{black}{There are still some caveats to this argument, however, 
since the scale of the end of inflation is not determined unless the details of the transition to radiation dominance are specified: we refer to \sect{conclusions} for a more complete discussion.}
\end{itemize}

We conclude this section by noting that the integral in \eq{Ngw-final} can be carried out analytically, giving (for $h^2_0\approx 0.5$, a pivot of $0.01\,\mpc^{-1}$, and taking $f_\tu{UV} = c k_\tu{end}/2\pi = \num{3.1d8}\,\hz$)
\begin{subequations}
\label{eq:Ngw-BBN-CMB}
\begin{align}
&\nbbn\approx\num{3d-6}\times\frac{r A_\tu{s}}{\nt}\bigg[\bigg(\frac{f}{f_\star}\bigg)^{\nt}\bigg]^{\num{3d8}\,\hz}_{\num{d-10}\,\hz}\,\,,
\notag \\
&\hphantom{\ncmb}\approx\num{3d-6}\times\frac{r A_\tu{s}}{\nt}\times(10^{25\nt} - 10^{7\nt})\,\,, \label{eq:Ngw-BBN} \\
&\ncmb\approx\num{3d-6}\times\frac{r A_\tu{s}}{\nt}\bigg[\bigg(\frac{f}{f_\star}\bigg)^{\nt}\bigg]^{\num{3d8}\,\hz}_{\num{d-15}\,\hz}\,\,,
\notag \\
&\hphantom{\nbbn}\approx\num{3d-6}\times\frac{r A_\tu{s}}{\nt}\times(10^{25\nt} - 63^{\nt})\,\,. \label{eq:Ngw-CMB}
\end{align}
\end{subequations}

\subsection{CMB distortions}
\label{sec:observables-distortions}

\noindent Once the tight-coupling approximation breaks down, the anisotropic stresses in the photon-baryon plasma become manifest, generating the dissipation of acoustic waves by Silk damping: at redshifts above $z \approx\num{2d6}\equiv z_{\mu,\text{i}}$, this energy released in the plasma is thermalized successfully by processes like elastic and double Compton scattering ($e^- + \gamma\to e^- + 2\gamma$), resulting again in a black-body spectrum with a higher temperature \cite{Jeong:2014gna}. At redshift between $z_{\mu,\text{i}}$ and $z\approx\num{5d4}\equiv z_{\mu,\text{f}}$, elastic Compton scattering allows to still achieve equilibrium, but photon number changing processes are frozen out due to the cosmic expansion and cannot re-establish a black-body spectrum. 
The result is a perturbed Planck spectrum that can be approximated by a Bose-Einstein distribution $1/(e^{x + \mu(x)} - 1)$ ($x\equiv h\nu/k_\tu{B}T$), 
where $\mu(x)$ can be identified as a chemical potential, 
independent on frequency away from the Rayleigh-Jeans tail \cite{Chluba:2011hw}.

The photon quadrupole anisotropy plays a crucial role in this dissipation process, giving rise to shear viscosity in the photon fluid \cite{Weinberg:1971mx, Kaiser:1983abc}. It is not, however, the only source of energy injection: the local quadrupole anisotropy is also sourced by tensor perturbations, without the need of photon diffusion \cite{Hu:1997hp}. Thomson scattering then mixes photons causing nearly scale independent dissipation \cite{Ota:2014hha, Chluba:2014qia}. 

Using the Bose-Einstein distribution $1/(e^{x + \mu(x)} - 1)$ plus the fact that, for $z_{\mu,\text{f}}\lesssim z\lesssim z_{\mu,\text{i}}$, the total number of photons is constant, one can show that for an amount of energy (density) $\delta E$ released into the plasma 
the resulting $\mu$-distortion is
\begin{equation}
\label{eq:mu}
\mu = 
\num{1.4}\int_{z_{\mu,\text{i}}}^{z_{\mu,\text{f}}}\dif z\frac{\dif(Q/\rho_\gamma)}{\dif z}\,\,,
\end{equation}
where $\dif(Q/\rho_\gamma)/\dif z$ is the energy injection as a function of redshift: $\dif(Q/\rho_\gamma)/\dif z$ will be also be a function of position, \ie there will be inhomogeneities in the chemical potential $\mu$.  
If we focus on the $\mu$-monopole $\braket{\mu}$, instead, one can show that: 
\begin{itemize}[leftmargin=*]
\item for scalars, 
$\braket{\dif(Q/\rho_\gamma)/\dif z}$ is
\begin{equation}
\label{eq:EI-scalar}
\begin{split}
\Braket{\frac{\dif(Q/\rho_\gamma)}{\dif z}}_\tu{s} &= \frac{9}{0.4 R_\nu + 1.5}\frac{\dif (k^{-2}_d)}{\dif z}\times \\
&\;\;\;\;\int_0^{+\infty}\frac{\dif^3 k}{4\pi}\frac{\Delta^2_\tu{s}(k)}{k}e^{-2k^2/k_d^2}\,\,,
\end{split}
\end{equation}
where $R_\nu = \rho_\nu/(\rho_\nu + \rho_\gamma)$ is approximately \num{0.41}. The damping wavenumber $k_\tu{d}$ is related to the mean squared diffusion distance $r_\tu{d}$ simply by $k_\tu{d} = \pi/r_\tu{d}$, while $r_\tu{d}$ is given by
\begin{equation}
\label{eq:diff-dist}
r_\tu{d}^2 = \pi^2\int_0^a\frac{\dif a'}{a'^3\sigma_\tu{T}n_e H}\bigg[\frac{R^2 + \frac{16}{15}(1+R)}{6(1+R)^2}\bigg]\,\,,
\end{equation}
with $\sigma_\tu{T}$ being the Thomson cross-section and $n_e$ the number density of free electrons;
\item for tensors, 
$\braket{\dif(Q/\rho_\gamma)/\dif z}$ is
\begin{equation}
\label{eq:EI-tensor}
\begin{split}
\Braket{\frac{\dif(Q/\rho_\gamma)}{\dif z}}_\tu{t} &= \frac{1.29}{48(1-R_\nu)}\times \\
&\;\;\;\;\int_0^{+\infty}\dif\log k\,\Delta^2_\tu{t}(k)\frac{\dif e^{-\Gamma\eta}}{\dif z}\,\,,
\end{split}
\end{equation}
where $\Gamma = 32 H (1 - R_\nu)/15 a \sigma_\tu{T}n_e$, the damping of the gravitational wave amplitude due to photons, is approximately equal to \num{5.9} during radiation domination, \ie when $\mu$-distortions are generated. The factor of \num{1.29/2} arises from the average $\braket{\mathcal{T}(x)}$ of the transfer function $\mathcal{T}(k\eta)$ for tensor perturbations \cite{footnote2}, once the effect of neutrino free-streaming \cite{Weinberg:2003ur, Watanabe:2006qe} is taken into account.
\end{itemize}
We note that in \eq{EI-tensor}, the integral in $\dif\log k$ extends on all wavenumbers from $k = 0$ to $k = +\infty$: at long wavelengths the time derivative of the transfer function for the gravitational wave amplitude  
vanishes, so that no super-horizon heating occurs (we will not reproduce the calculations here, and refer to \cite{Watanabe:2006qe} for details). At small scales, however, there is an UV cutoff given by the photon mean free path $k_\tu{mfp} = a \sigma_\tu{T}n_e \approx\num{4.5d-7}(1+z)^2\,\mpc^{-1}$, since for larger momenta photons stream quasi-freely and add little heating \cite{Chluba:2014qia}.

Taking into account the cutoff for the tensor case, one can obtain estimates for the contribution of scalar and tensor perturbations to the $\mu$-distortion monopole 
\begin{subequations}
\label{eq:mu-int}
\begin{align}
&\braket{\mu}_\tu{s}\approx\num{2.3}\times\frac{A_\tu{s}}{\ns - 1}\bigg[\bigg(\frac{k}{k_\star}\bigg)^{\ns - 1}\bigg]_{k_\tu{d}(z_{\mu,\text{f}})}^{k_\tu{d}(z_{\mu,\text{i}})}\,\,, \label{eq:mu-int-s} \\
&\braket{\mu}_\tu{t}\approx\num{7.3d-6}\times\frac{r A_\tu{s}}{\nt}\bigg[\bigg(\frac{k}{k_\star}\bigg)^{\nt}\bigg]_{k_\tu{mfp}(z_{\mu,\text{f}})}^{k_\tu{mfp}(z_{\mu,\text{i}})}\,\,. \label{eq:mu-int-t}
\end{align}
\end{subequations}
where $1/k_\tu{d}(z)$ ($1/k_\tu{mfp}(z)$) is the Silk damping scale (photon mean free path) at redshift $z$. Plugging in numerical values, we get
\begin{subequations}
\label{eq:mu-int-num}
\begin{align}
&\braket{\mu}_\tu{s}\approx\num{2.3}\times\frac{A_\tu{s}}{\ns - 1}\bigg[\bigg(\frac{k}{k_\star}\bigg)^{\ns - 1}\bigg]_{\num{46}\,\mpc^{-1}}^{\num{1.1d4}\,\mpc^{-1}}\,\,, \label{eq:mu-int-s-num} \\
&\braket{\mu}_\tu{t}\approx\num{7.3d-6}\times\frac{r A_\tu{s}}{\nt}\bigg[\bigg(\frac{k}{k_\star}\bigg)^{\nt}\bigg]_{\num{1.1d3}\,\mpc^{-1}}^{\num{1.8d6}\,\mpc^{-1}}\,\,. \label{eq:mu-int-t-num}
\end{align}
\end{subequations}

Tensors and scalar modes are not the only source of distortions: a third source is the so-called adiabatic cooling of photons \cite{Chluba:2011hw, Khatri:2011aj}. The difference in adiabatic indices of photons and baryons implies that, in the tight-coupling era, the baryonic matter must continuously extract energy from the CMB in order to establish $T_\tu{b}\sim T_\gamma$. This cooling of the Planck spectrum would, in principle, lead to a Bose-Einstein condensation: 
however the time-scale for this to happen is quite long and no condensate is in reality possible.  
This effect can be described by a negative contribution to the overall $\dif(Q/\rho_\gamma)/\dif z$, given by \cite{footnote3}
\begin{equation}
\label{eq:EI-ac}
\Braket{\frac{\dif(Q/\rho_\gamma)}{\dif z}}_\tu{ac} = 
\frac{3k_\tu{B}[2n_\tu{H}(z) + 3n_\tu{He}(z)]}{2 a_R(1+z)T^3_\gamma}\,\,,
\end{equation}
where $a_\tu{R}$ is the radiation constant.

We take also this effect (which gives 
a $\braket{\mu}_\tu{ac}$ of order \num{-2.8d-9}) into account in our analysis: 
the total $\mu$-distortion is, then, the sum $\braket{\mu} = \braket{\mu}_\tu{s} + \braket{\mu}_\tu{t} + \braket{\mu}_\tu{ac}$.

\subsection{Pulsar timing + ground- and space-based interferometers}
\label{sec:observables-pulsar+interferometers}

\noindent In this short section we briefly review the physics of interferometers and pulsar timing. We refer to \cite{Maggiore:1999vm} (and references therein) for a more detailed 
treatment, which is outside of the scope of this paper.

\subsubsection{Pulsar timing}
\label{sec:observables-pulsar}

\noindent Pulsars are neutron stars formed during the supernova explosion of stars with \num{5} to \num{10} solar masses. 
Because of the great intrinsic stability of their pulsation periods, precision timing observations of pulsars (in particular, millisecond pulsars), can be used to detect GWs propagating in our Galaxy \cite{Detweiler:1979abc}: the reason is that the observed pulse frequencies will be modulated by gravitational waves passing 
between the pulsar and the Earth. This will give rise to a \emph{timing residual}: 
the deviation of the observed pulse time of arrival from what is expected given our knowledge of the motion of the pulsar and the strict periodicity of the pulses. If one considers a 
gravitational wave propagating towards the Earth and traveling in the $z$-axis, the GW-induced timing residual of an observation at time $t$ (calling $t_0$ the starting time of the experiment) is given by \cite{Maggiore:1999vm}
\begin{equation}
\label{eq:time-delay}
\begin{split}
\delta t^\tu{GW} = \int_{t_0}^{t}\dif t\frac{1-\cos\theta}{2}[&\cos 2\psi h_+(t - s/c) \\
&+ \sin2\psi h_\times(t - s/c)]\,\,,
\end{split}
\end{equation}
where $\theta$ is the polar angle of the Earth-pulsar direction measured from the $z$-axis, $\psi$ is a rotation in the plane orthogonal to the direction of propagation of the wave (in this case the $(x,y)$ plane), corresponding to a choice of the axes to which the $+$ and $\times$ polarization are referred, and $t - s/c$ is the time when the GW crossed the Earth-pulsar direction (with $s$ being the distance along the path). 

One can, therefore, relate the (
variance of) these timing residuals to the energy density of the GW background. More precisely, 
EPTA will be sensitive to the integral of the spectrum $\Omega_\tu{GW}(f)$ in a small interval of frequencies 
around the frequency $f_\tu{PTA}$, which 
will be in the range $\sim\num{d-9}\,\hz\,\div\,\num{d-7}\,\hz$.

\subsubsection{Ground- and space-based interferometers}
\label{sec:observables-interferometers}

\noindent The basis of present detectors is the effect of gravitational waves on the separation of adjacent masses on Earth or in space \cite{Pitkin:2011yk, Adhikari:2013kya}. GW strength is characterized by the change $2\Delta L/L$ in the separation of two masses a distance $L$ apart. We consider a wave propagating in the $z$-direction, with \emph{strain} $h_{\mu\nu}$ given by
\begin{equation}
\label{eq:h-mu-nu}
h_{\mu\nu} =
\begin{pmatrix}
0 & 0 & 0 & 0 \\
0 & -h_+ & h_\times & 0 \\
0 & h_\times & h_+ & 0 \\
0 & 0 & 0 & 0
\end{pmatrix}
\,\,,
\end{equation}
and place two masses at the origin and at $x = L$. We can measure the distance between the two masses by sending a laser beam from the source at $x = 0$ to $x = L$ and back, and measure the phase of the returned beam relative to that of the source. In the presence of a gravitational wave, the ``round-trip (rt) phase'' $\Phi(t_\tu{rt})$ is
\begin{equation}
\label{eq:phi-x}
\Phi_x(t_\tu{rt}) = 2\frac{2\pi\nu}{c}\int_0^L\dif x\,\sqrt{g_{xx}}\approx 2\bigg(1 - \frac{h_+}{2}\bigg)\frac{2\pi L}{\lambda}\,\,,
\end{equation}
for ``plus'' oriented wave with a period much longer than the round-trip light travel time. With an analogous set-up along the $y$-axis, instead, one gets
\begin{equation}
\label{eq:phi-y}
\Phi_y(t_\tu{rt})\approx 2\bigg(1 + \frac{h_+}{2}\bigg)\frac{2\pi L}{\lambda}\,\,.
\end{equation}
The difference in phase shift $\Delta\Phi = \Phi_y(t_\tu{rt}) - \Phi_x(t_\tu{rt})$ between the two arms will be proportional, then, to the gravitational wave amplitude $h_+$. 
Considering averages of the squared difference in phase shift, direct detection experiments are able to probe the spectrum of a stochastic background of gravitational waves.

\section{Method}
\label{sec:method}

\subsection{Monte Carlo method and datasets}
\label{sec:method-MCMC}

\noindent We perform a Monte Carlo Markov Chain (MCMC) analysis, using the publicly available code \texttt{cosmomc}~\cite{Lewis:2002ah, Lewis:2013hha}. 
We vary the six standard $\Lambda$CDM cosmological parameters, namely the baryon density $\Omega_b h^2$, the cold dark matter density $\Omega_c h^2$, the sound horizon angular scale $\theta$, the reionization optical depth $\tau$, the amplitude $\log (10^{10} A_\tu{s})$ and spectral index $\ns$ of the primordial 
spectrum of scalar perturbations. 
We add to these the tensor-to-scalar ratio $r$ and the tensor spectral index $\nt$. 
Unless otherwise stated, we normalize the inflationary parameters to the pivot wavelength $k = 0.01\,\mpc^{-1}$, roughly corresponding to $\ell\simeq 150$, using the approximate formula $\ell\sim1.35\times 10^4 (k/\mpc^{-1})$. This is where the data published by the BKP collaboration are most sensitive and it is close to the decorrelation scale between the tensor amplitude and slope for Planck and BKP joint constraints \cite{Ade:2015lrj}. 

Our reference dataset is based on CMB temperature and polarization anisotropies. We analyze the $BB$ power spectrum as measured by the BKP collaboration (first five bandpowers)~\cite{Ade:2015tva}, in combination with the temperature and polarization \emph{Planck} likelihood \cite{Aghanim:2015xee}. More precisely, we make use of the $TT$, $TE$, $EE$ high-$\ell$ likelihood together with the $TQU$ pixel-based low-$\ell$ likelihood. \tblack{We also compare the BKP $BB$ power spectrum to the recently released BICEP2/\emph{Keck Array} polarization data (BK14, hereafter) \cite{Array:2015xqh}.}
Note that, when using 
these datasets, we perform the analysis both with 
and without $\ngw$. The second case could seem unrealistic: in fact, if GWs are present, they will surely contribute to the number of effective radiation d.o.f. 
However, it is possible to have scenarios 
where the contribution of neutrinos to $\neff$ is lower than \num{3.046} \cite{Kawasaki:1999na, Kawasaki:2000en, deSalas:2015glj}. For this reason instead of referring to a specific non-standard model for computing the contribution of neutrinos to $\neff$ and subsequently add the $\ngw$ contribution, we follow a more generic approach and we assume a total value of relativistic degrees of freedom of $3.046$ (in agreement with current cosmological constraints) as the final sum of the two contributions.

We will consider, then, the following extensions to our reference dataset:
\begin{itemize}[leftmargin=*]
\item BAO and Deuterium. We use baryon acoustic oscillations (BAO) to break geometrical degeneracies, as reported in \cite{Planck:2015xua}: the surveys included are 6dFGS \cite{Beutler:2011hx}, SDSS-MGS \cite{Ross:2014qpa}, BOSS LOWZ \cite{Anderson:2013zyy} and CMASS-DR11 \cite{Anderson:2013zyy}. We use primordial Deuterium abundance measurements \cite{Cooke:2013cba} to constrain the relativistic number of degrees of freedom, assuming standard BBN. 
We do not use primordial helium measurements, 
since they are less constraining: this is due to the uncertainty in the neutron lifetime, which 
affects the computation of the helium abundance that will be then compared to observations 
(for a recent discussion about the effect of this uncertainty on cosmological parameter estimation, we refer to \cite{Salvati:2015wxa}). When used in combination with CMB data, we will denote this joint dataset by 
EXT;
\item FIRAS limits on deviations of the CMB from a blackbody spectrum, $\mu = (1 \pm 4)\times 10^{-5}$ (at $\limit{68}$) \cite{Fixsen:1996nj};
\item LIGO-Virgo limit on $\Omega_\tu{GW}(f)$ for frequencies in the band $\num{41.5}\,\hz\,\text{--}\,\num{169.25}\,\hz$ coming from the LIGO and Virgo joint analysis \cite{Aasi:2014zwg}, \ie $\Omega_\tu{GW}(f)\leq\num{5.6d-6}$ ($\limit{95}$). 
We note that the scales at which LIGO-Virgo is sensitive are likely dominated by astrophysical GW backgrounds (such as, {\it e.g.}, gravitational waves from binary mergers or rotating neutron stars): for this reason the limits that we will obtain on $\nt$ in this case must be regarded as conservative;
\item pulsar constraints on a stochastic relic GW background at $f = \num{2.8d-9}\,\hz$ obtained by EPTA in \cite{Lentati:2015qwp}, \ie $\Omega_\tu{GW}(f)\leq\num{1.2d-9}$ ($\limit{95}$).
\end{itemize}


\subsection{Simulated datasets for forecasts}
\label{sec:method-fiducials+specifics}

\noindent For our forecasts we generate simulated datasets following the 
approach described in \cite{Lewis:2005tp} (see also \sect{appendix-forecasting}). Regarding the presence of foregrounds (like dust and synchrotron emission), we assume that these can be 
removed 
after 
being characterized by the high- and low-frequency channels of COrE (\ie $\nu\gtrsim 100\,\mathrm{GHz}$ and $\nu\lesssim 230\,\mathrm{GHz}$), where the CMB is subdominant. We also assume that uncertainties due to foreground removal are smaller than instrumental noise (which we take as white and isotropic), and that beam uncertainties are negligible. We account for the impossibility of observing the full sky simply by reducing the degrees of freedom at each multipole $\ell$ with the sky fraction $f_\mathrm{sky}$ (see \sect{appendix-forecasting}), and ignoring the induced correlations between different $\ell$. Our fiducial model is described in \tab{fiducials}, while the specifics used for the COrE-like mission are listed in \tab{COrE-specifics}.

\textcolor{black}{In some cases, we consider also the upgraded version of current interferometer experiments, in combination with COrE simulated data. 
In particular, we focus our attention on AdvLIGO \cite{TheLIGOScientific:2014jea} (which we refer to as aLIGO in our plots), that will be able to reach a sensitivity of $10^{-9}$ on $\Omega_\tu{GW}(f)$ at 
$f = 100\,\hz$.}

\textcolor{black}{We also compare the forecasts from COrE alone with those from the combination \planckBKP~+~AdvLIGO \tblack{and \emph{Planck} + BK14 + AdvLIGO}: since 
a COrE-like mission (\ie COrE++, see \cite{core++}) is 
still in proposal stage, it is worth to investigate how limits on tensor parameters will improve 
thanks only to advancements in GW direct detection.} 

\begin{table}[!hbtp]
\begin{center}
\begin{tabular}{lc}
\toprule
\horsp
Parameter \vertsp Fiducial value \\
\hline
\horsp
$\Omega_\mathrm{b} h^2$   \vertsp  $0.02225$ \\
\horsp
$\Omega_\mathrm{c} h^2$  \vertsp $0.1198$ \\
\horsp
$100\theta_\mathrm{MC}$  \vertsp  $1.04077$ \\
\horsp
$\tau$  \vertsp $0.079$ \\
\horsp
$\ns$   \vertsp  $0.9645$ \\
\horsp
$\log(10^{10} A_\tu{s})$  \vertsp $3.145$ \\
\horsp
$H_0$  \vertsp $67.77$ \\
\horsp
$z_\mathrm{eq}$ \vertsp $3394$ \\
\botrule
\end{tabular}
\caption{\footnotesize{Fiducial cosmological parameters used for the COrE forecasts: the fiducials for the tensor parameters $r$ and $\nt$ are specified in the main text.}}
\label{tab:fiducials}
\end{center}
\end{table}

\begin{table}[!hbtp]
\begin{center}
\begin{tabular}{cccc}
\toprule
\horsp
channel \vertsp FWHM \vertsp $w^{-1/2}$ --- $T$ \vertsp $w^{-1/2}$ --- $Q$, $U$ \\
($\text{GHz}$) \vertsp ($\text{arcmin}$) \vertsp ($\mu\text{K}\cdot\text{arcmin}$) \vertsp ($\mu\text{K}\cdot\text{arcmin}$) \\
\hline
\horsp
$105$ \vertsp $10$ \vertsp $2.68$ \vertsp $4.63$ \\
\horsp
$135$ \vertsp $7.8$ \vertsp $2.63$ \vertsp $4.55$ \\
\horsp
$165$ \vertsp $6.4$ \vertsp $2.67$ \vertsp $4.61$ \\
\horsp
$195$ \vertsp $5.4$ \vertsp $2.63$ \vertsp $4.54$ \\
\horsp
$225$ \vertsp $4.7$ \vertsp $2.64$ \vertsp $4.57$ \\
\botrule
\end{tabular}
\caption{\footnotesize{Temperature and polarization noise (in $\mu\mathrm{K} \cdot \text{arcmin}$) and beam (FWHM in $\text{arcmin}$) specifications of the COrE experiment, from \cite{Bouchet:2011ck}. We suppose that the ten additional frequency channels (from $45\,\text{GHz}$ to $795\,\text{GHz}$) are used for foreground cleaning. The fraction of sky covered is $\fsky = 0.8$.}}
\label{tab:COrE-specifics}
\end{center}
\end{table}

\subsection{Delensing}
\label{sec:method-delensing}

\noindent Tensor modes are not the only source of $B$-mode polarization: gravitational lensing, in fact, generates a non-Gaussian $B$-mode signal \cite{Zaldarriaga:1998ar}. While interesting in its own right (see \cite{Smith:2008an} for a review), it acts as another source of foregrounds when the goal is studying primordial tensor modes. 
However, if the lensing potential is reconstructed on small scales, one can remove the lensing contribution to 
the CMB $B$-mode spectrum at large 
scales, where the contribution from tensors dominates \cite{Knox:2002pe, Kesden:2002ku, Seljak:2003pn, Smith:2010gu}.

\begin{figure}[!hbtp]
\includegraphics[width=0.48\textwidth]{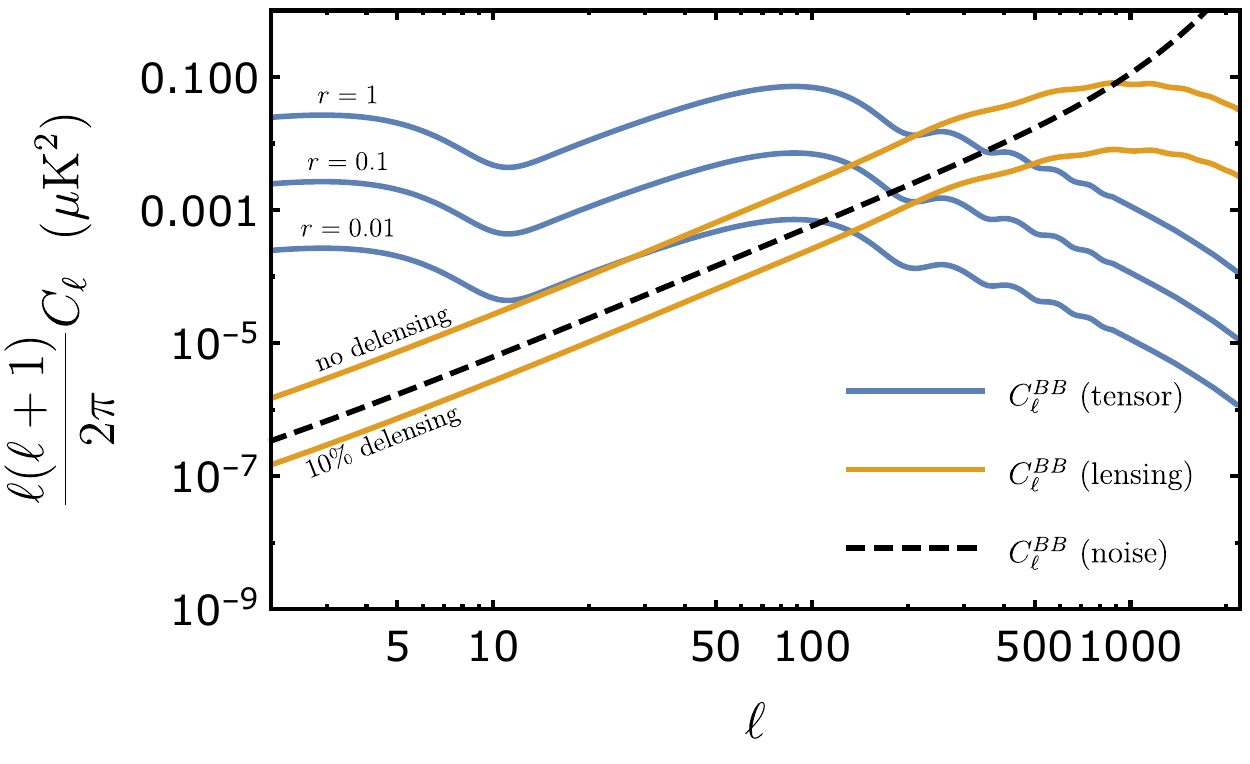}
\caption{\footnotesize{Tensor and lensing $B$-modes, together with noise bias 
for the COrE experiment. The blue spectra correspond 
$r = 1$, $0.1$, $0.01$ and $\nt = 0$, while 
the orange ones are the lensing $B$-mode spectrum 
with and without a rescaling by a factor of \num{0.1}. We see that in the case of a $10\%$ delensing the $C^\tu{lens}_\ell$ go completely below noise level.}}
\label{fig:delensing}
\end{figure}

In the recent work \cite{Errard:2015cxa} (more precisely, we refer to Fig.~4 -- left panel) 
it is shown that, for an experiment with a noise level of order $\sim 1\,\mu\K\cdot\mathrm{arcmin}$ post component separation, one can bring the power of lensing $B$-modes down to $10\%$ of their original value (see \fig{delensing}). Since we expect that COrE could reach these noise levels after component separation has been carried out, we implement a $10\%$ delensing in our forecasts 
following \cite{Creminelli:2015oda}, \ie by rescaling the lensing $B$-mode angular spectrum $C^\tu{lens}_\ell$ of a factor of \num{0.1}.
This removal of lensing $B$-modes allows us to gain sensitivity to the scale dependence of $\Delta^2_\tu{t}$ on a 
wider range of multipoles. In \fig{delensing} we see that if the $C^\tu{lens}_\ell$ are reduced to $10\%$ in power, for $r = 0.1$ there are $\sim 100$ more multipoles available before the noise becomes larger than the signal. 
For lower values of the tensor-to-scalar ratio the gain would be even larger \cite{footnote4}.

\section{Results}
\label{sec:results}

\subsection{Current data}
\label{sec:results-current}

\begin{figure*}
\begin{center}
\begin{tabular}{c c}
\includegraphics[width=\columnwidth]{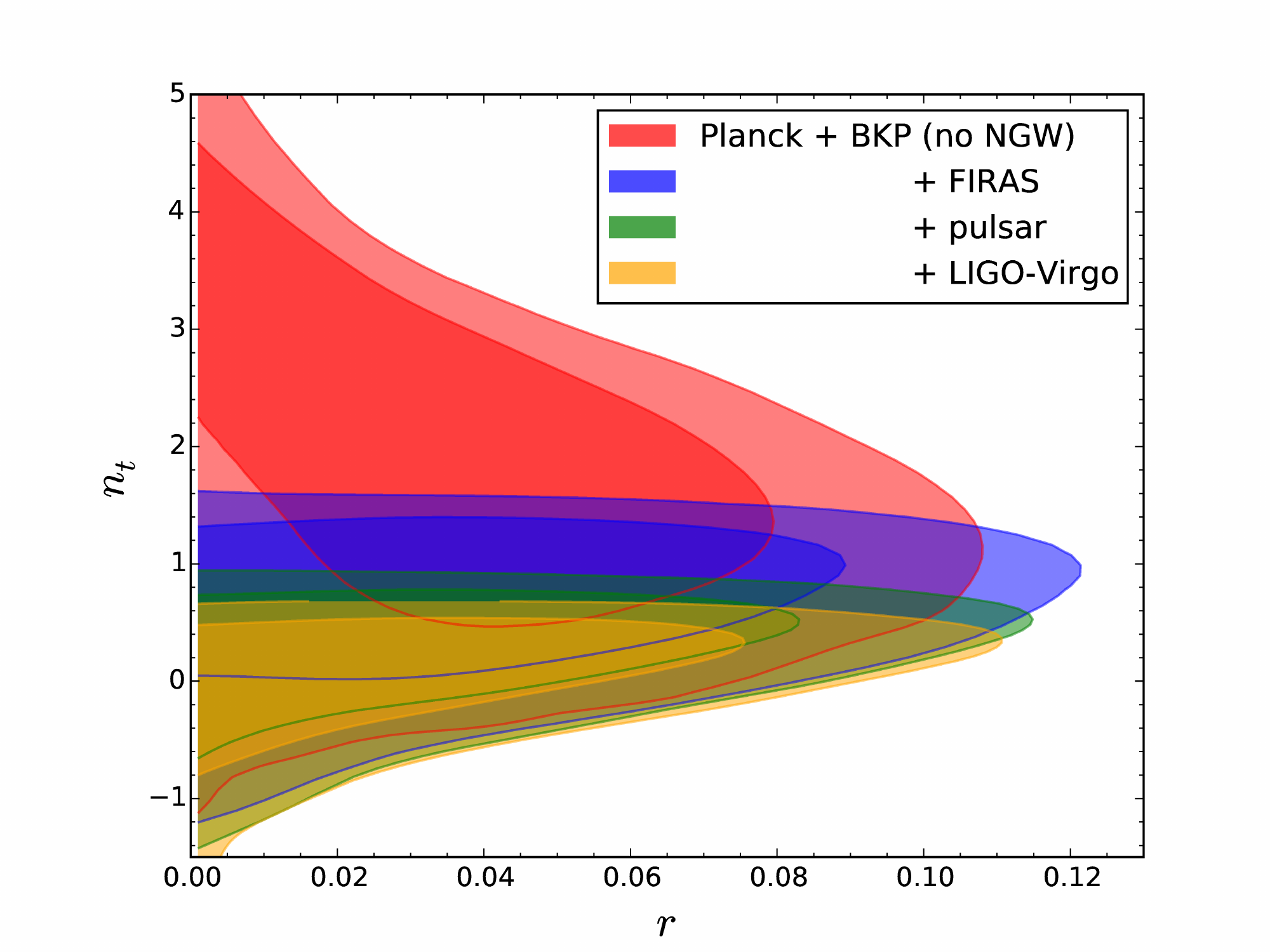}
&\includegraphics[width=\columnwidth]{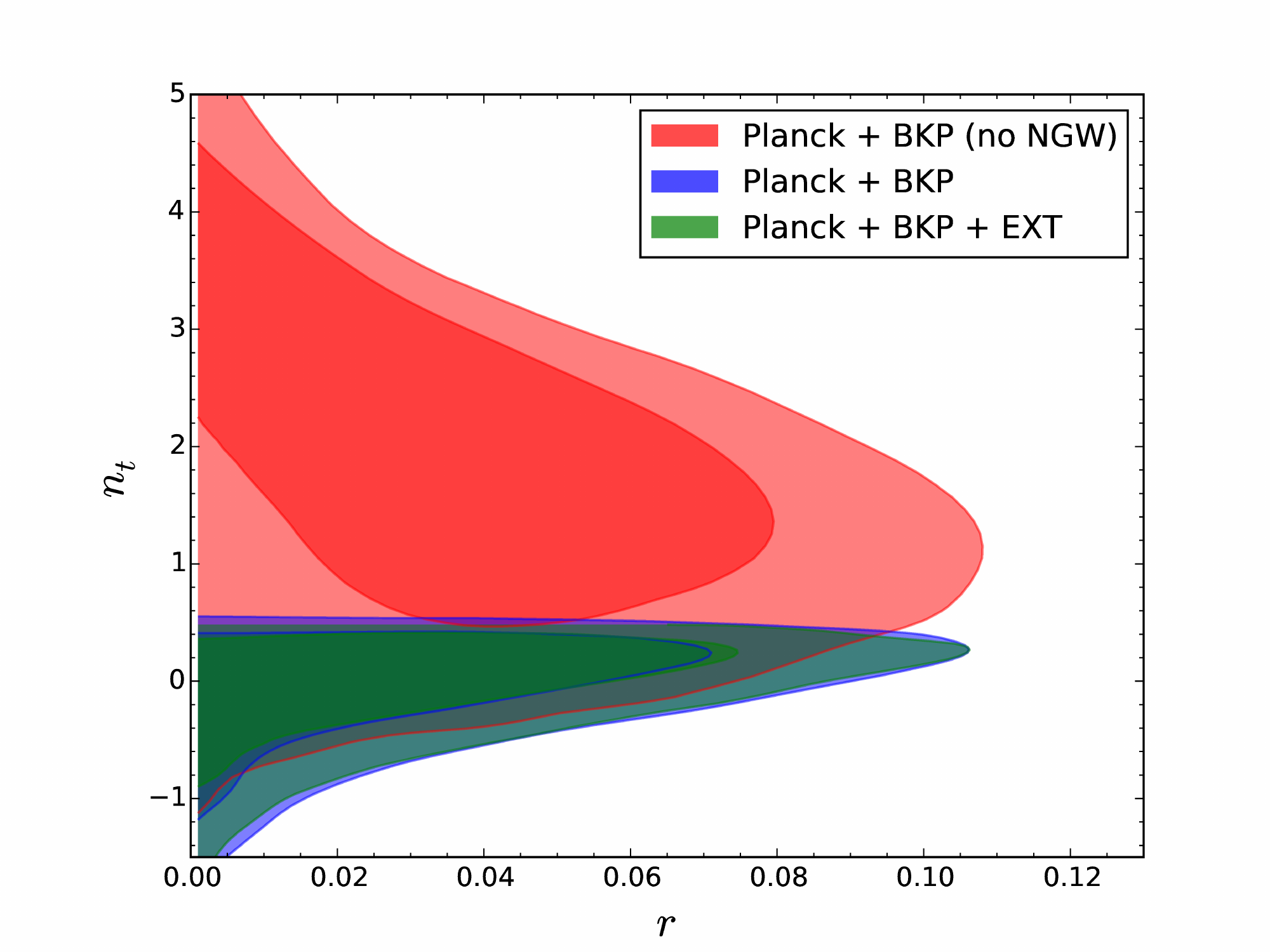}
\end{tabular}
\end{center}
\caption{\footnotesize{Two-dimensional 
posterior distributions for $r$ and $\nt$: the contours in the left (right) panel are obtained without (with) the inclusion of $\ngw$. In both panels the red contour is the result for the ``vanilla'' $\Lambda\text{CDM} + r + \nt$ model, using the \planckBKP~dataset. The corresponding $\limit{95}$ results for $r$ and $\nt$ are reported in \tab{results-current}.}}
\label{fig:double_plot_r_nt}
\end{figure*}

\begin{figure*}
\begin{center}
\begin{tabular}{c c}
\includegraphics[width=\columnwidth]{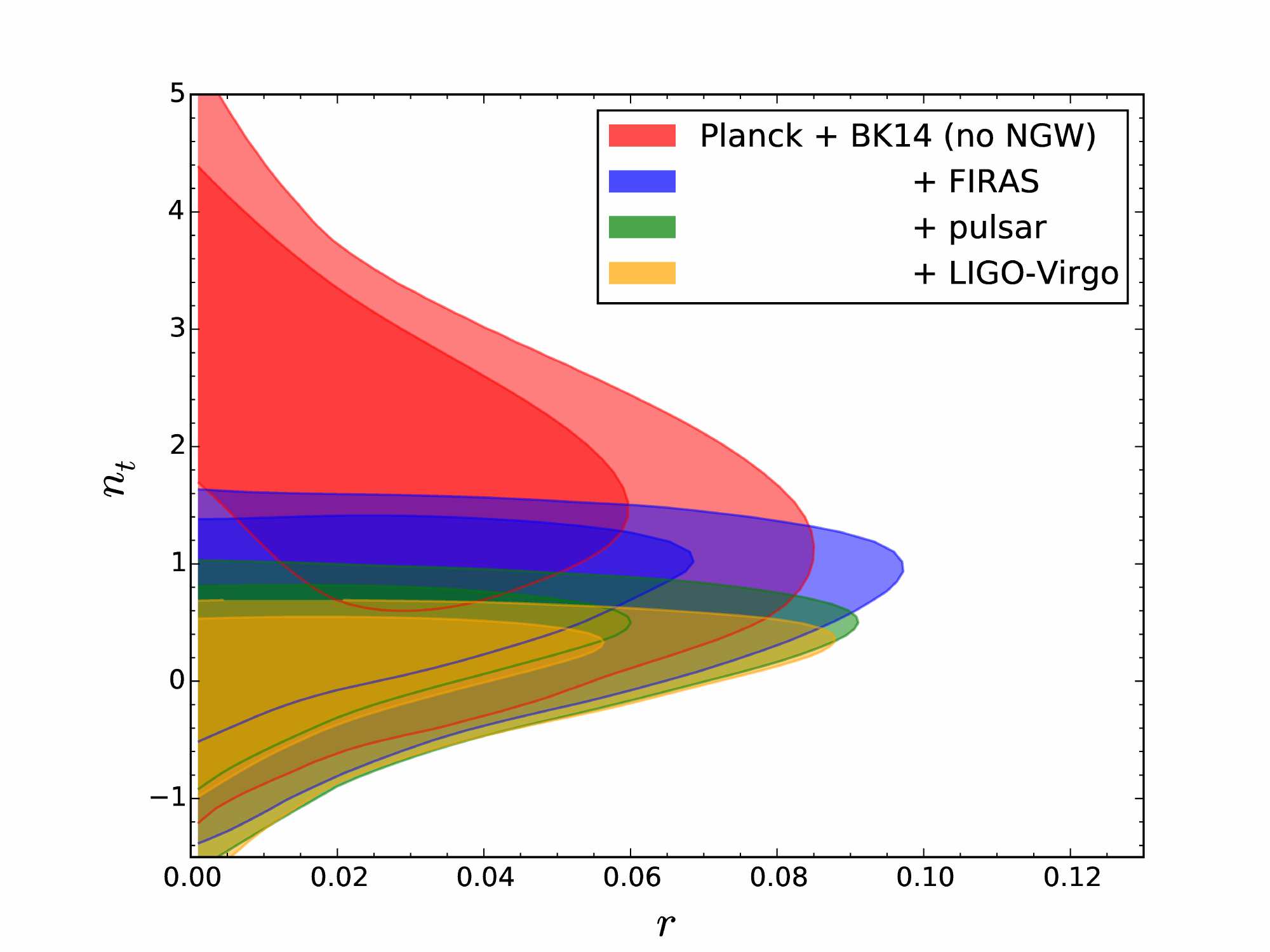}
&\includegraphics[width=\columnwidth]{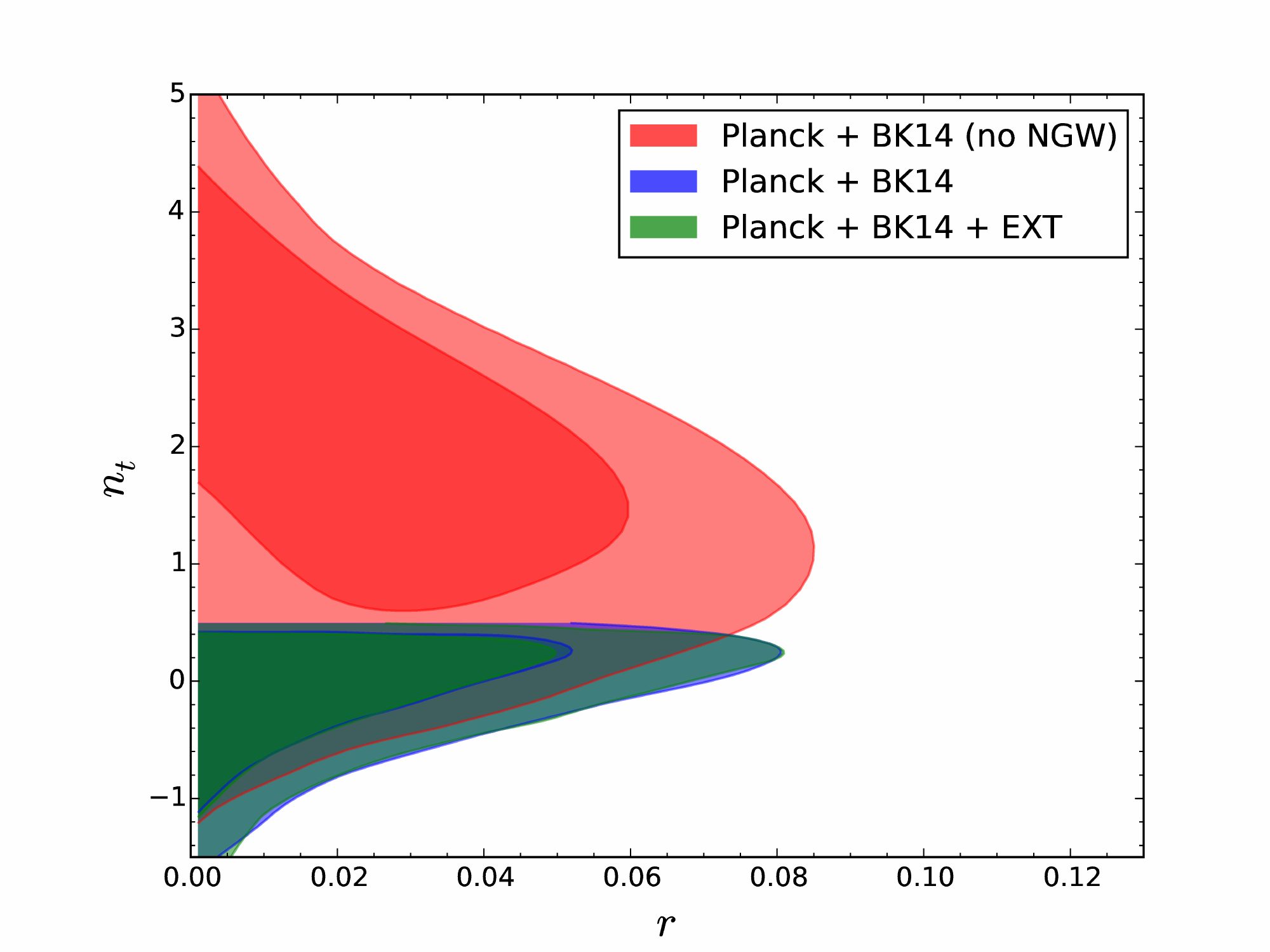}
\end{tabular}
\end{center}
\caption{\footnotesize{Two-dimensional 
posterior distributions for $r$ and $\nt$, using BK14 polarization data: the contours in the left (right) panel are obtained without (with) the inclusion of $\ngw$. 
The corresponding $\limit{95}$ results for $r$ and $\nt$ are reported in \tab{results-current-BK}.}}
\label{fig:double_plot_r_nt-BK}
\end{figure*}

\noindent We begin this section with a discussion about priors. In the absence of primordial GWs detection, the bounds 
on the tensor tilt will be prior dependent: while taking a flat prior around $\nt = 0$ is strongly motivated by scale invariance, there is no equivalently strong guidance on the prior range: in fact the effect of a very blue tilt could always be 
cancelled by a very small tensor-to-scalar ratio. For this reason, the limits we put on $\nt$ depend on which sampling of $r$ we use for our MCMC exploration of parameter space: 
we have used in our analysis a linear prior on $r$, with $r > 0.001$ \cite{footnote-prior}. 
Had we chosen, \textit{e.g.}, a logarithmic prior on $r$, the tails of the two-dimensional contours depicted in 
Figs.~\ref{fig:double_plot_r_nt}, \ref{fig:double_plot_r_nt-BK}  would have extended on a wider range on the $\nt$-axis, and the marginalized constraints on the tilt would have 
degraded: the more one samples regions at low $r$, the less tight the bounds on the tilt will be. 
This is a very important point, that must be kept in mind when interpreting Figs.~\ref{fig:double_plot_r_nt}, \ref{fig:double_plot_r_nt-BK} and Tabs.~\ref{tab:results-current}, \ref{tab:results-current-BK}. 

Turning to the actual results, the first thing we notice 
from the left panel of \fig{double_plot_r_nt} is that the posteriors for $\nt$ favor a blue tilt, when $\ngw$ is turned off and no additional observables besides CMB anisotropies are considered. This is due to the fact that we normalize our spectra at a pivot of $0.01\,\mpc^{-1}$. In fact, in order to be consistent with the low tensor power at large scales, where the constraints from the \planck~data come from, a blue tilt is needed \cite{footnote5}.

When we add the information from spectral distortions, pulsar timing or GW direct detection, instead, we find that
the upper limits for the tilt decrease, 
in agreement 
with the fact that a too large $\nt$ would lead to a large and detectable tensor signal at small scales. Moreover, there is an almost horizontal cut in the two-dimensional posteriors for $r$ and $\nt$: the reason is that the dependence of these three 
observables on the tilt is exponential, and this causes the posterior probability density function to be very steep in the $\nt$ direction. 
\tblack{The left panel of \fig{double_plot_r_nt} shows that the most constraining dataset is \planckBKPLIGO, followed by \planckBKPpulsar~and \planckBKPFIRAS.}

This 
hierarchy is expected 
since the tensor and scalar contributions for spectral distortions 
are degenerate (as 
one can see from \sect{observables-distortions}), \textcolor{black}{and Eqs. \eqref{eq:mu-int-num} show that the scalar contribution $\braket{\mu}_\tu{s}$ dominates over the tensor one unless $\nt$ is very large}. Regarding the \planckBKPpulsar~and \planckBKPLIGO~datasets we could have expected to obtain better constraints with pulsar timing than with direct GW measurements, since the former put more stringent upper limits on $\Omega_\tu{GW}$.
The reason why this does not happen is that the frequency range  
where pulsar timing operates is closer to the horizon size at recombination than LIGO-Virgo frequencies, \textcolor{black}{therefore giving a weaker lever arm to estimate the scale dependence of the primordial tensor spectrum}. 

Our best 
bounds on the tensor parameters, obtained using the \planckBKPLIGO~dataset, are 
$r<0.085$ and $\nt = 0.04 _{-0.85}^{+0.61}$ (both at $\limit{95}$). 

Another thing that we notice is the following: since the low tensor power at large scales cannot be anymore accommodated by having a blue tensor tilt, 
the upper limits on the tensor-to-scalar ratio decrease as those on $\nt$ become tighter. This does not happen, however, for the \planckBKPFIRAS~dataset: \textcolor{black}{the reason is that 
the best-fit of \planckBKP~is 
excluded by the combination \planckBKPFIRAS, so regions of parameter space that were before 
forbidden at more than $2\sigma$ become again compatible with data at $\limit{95}$. 
The same argument applies also to \planckBKPpulsar~and \planckBKPLIGO: in that case, however, the constraints on the tilt derived from $\Omega_\tu{GW}$ are strong enough that $r$ must be brought down in order to have consistency with the \planckBKP~bounds on the large-scale tensor power.} 

\tblack{The left panel of \fig{double_plot_r_nt-BK} and \tab{results-current-BK} show that switching from BKP to BK14 polarization data has mainly the effect of tightening the bounds on the tensor-to-scalar ratio, while those for $\nt$ are practically unaffected.  
Given that the BK14 dataset puts more precise bounds on the $BB$ power spectrum, one could expect to obtain also strongest constraints on the tensor tilt. However, since BK14 spectra prefer values of $r$ lower than BKP ones, 
the gain from the higher experimental accuracy is cancelled by the lost sensitivity of the angular spectra to variations in $\nt$.}

\tblack{Similarly to the previous case, the best bounds on tensor parameters ($r < 0.067$, $\nt = 0.00_{-0.91}^{+0.68}$, both at $\limit{95}$) are obtained by the combination of CMB anisotropies and direct detection experiments.}

When we add the contribution $\ngw$ to the effective number of degrees of freedom $\neff$, \tab{results-current} shows that we obtain 
more stringent constraints on $r$ and $\nt$, while we see from \tblack{the right panel of} \fig{double_plot_r_nt} that the steep slope of the posterior in the $\nt$ direction is reproduced (recall Eqs.~\eqref{eq:Ngw-BBN-CMB}). In particular 
we see that in this case, even if we are using ``just CMB 
information'' (\ie the effect of $\neff$ on CMB anisotropies only), we reach a constraining power comparable to or even better 
than CMB 
combined with 
GW direct detection experiments. Of course, by adding external astrophysical datasets (as BAO and primordial Deuterium abundance) we obtain even tighter bounds. 
Our best 
limits, obtained using \planckBKPEXT, are $r<0.080$ and $\nt = -0.05_{-0.80}^{+0.57}$, both at $\limit{95}$.

\tblack{Also in this case, adding the BK14 dataset leads to better constraints on $r$: we see from \tab{results-current-BK} that considering the \emph{Planck} + BK14 + EXT dataset we reach $r < 0.061$ ($\limit{95}$).}

\begin{table}[!hbtp]
\begin{center}
\begin{tabular}{lcc}
\toprule
\horsp
Dataset \vertsp $r$ \vertsp $\nt$ \\
\hline
\horsp
\planckBKP \vertsp $<0.089$\vertsp $1.7_{-2.0}^{+2.2}$ \\
\horsp
\planckBKPFIRAS \vertsp$<0.098$ \vertsp$0.65_{-1.1}^{+0.86}$ \\
\horsp
\planckBKPpulsar \vertsp$<0.088$ \vertsp$0.20_{-0.96}^{+0.69}$ \\ 
\horsp
\planckBKPLIGO \vertsp $<0.085$ \vertsp$0.04_{-0.85}^{+0.61}$ \\ 
\hline
\horsp
\planckBKP, with $\ngw$ \vertsp $<0.082$ \vertsp $-0.05_{-0.87}^{+0.58}$ \\
\horsp
\planckBKPEXT, with $\ngw$ \vertsp $<0.080$ \vertsp $-0.05_{-0.80}^{+0.57}$ \\ 
\hline
\hline
\horsp
\planckBKP~+ aLIGO \vertsp $< 0.078$ \vertsp $-0.09_{-0.78}^{+0.54}$ \\
\botrule
\end{tabular}
\caption{\footnotesize{Constraints at $\limit{95}$ on the tensor-to-scalar ratio $r$ and the tensor spectral index $\nt$ for the listed datasets: the first four results are obtained without considering the GW contribution to $\neff$. For a detailed description of the datasets used in the analysis see \sect{method-MCMC}. For the \planckBKP~+ aLIGO forecast we assumed no detection for AdvLIGO.}}
\label{tab:results-current}
\end{center}
\end{table}

\begin{table}[!hbtp]
\begin{center}
\begin{tabular}{lcc}
\toprule
\horsp
Dataset \vertsp $r$ \vertsp $\nt$ \\
\hline
\horsp
\emph{Planck} + BK14 \vertsp $<0.067$\vertsp $1.8_{-2.1}^{+2.0}$ \\
\horsp
\emph{Planck} + BK14 + FIRAS \vertsp$<0.078$ \vertsp$0.63_{-1.16}^{+0.89}$ \\
\horsp
\emph{Planck} + BK14 + pulsar \vertsp$<0.070$ \vertsp$0.17_{-1.03}^{+0.75}$ \\ 
\horsp
\emph{Planck} + BK14 + LIGO-Virgo \vertsp $<0.067$ \vertsp$0.00_{-0.91}^{+0.68}$ \\ 
\hline
\horsp
\emph{Planck} + BK14, with $\ngw$ \vertsp $<0.061$ \vertsp $-0.12_{-0.84}^{+0.65}$ \\
\horsp
\emph{Planck} + BK14 + EXT, with $\ngw$ \vertsp $<0.061$ \vertsp $-0.10_{-0.88}^{+0.63}$ \\ 
\hline
\hline
\horsp
\emph{Planck} + BK14~+ aLIGO \vertsp $<0.060$ \vertsp $-0.16_{-0.88}^{+0.63}$ \\ 
\botrule
\end{tabular}
\caption{\footnotesize{Same as \tab{results-current}, but considering BK14 polarization data in addition to \emph{Planck} power spectra. For a detailed description of the datasets we refer to \sect{method-MCMC}. As in table \ref{tab:results-current} for AdvLIGO we assumed no detection of primordial GWs.}}
\label{tab:results-current-BK}
\end{center}
\end{table}

\subsection{Forecasts}
\label{sec:results-forecasts}


\begin{figure*}
\begin{center}
\begin{tabular}{c c}
\includegraphics[width=\columnwidth]{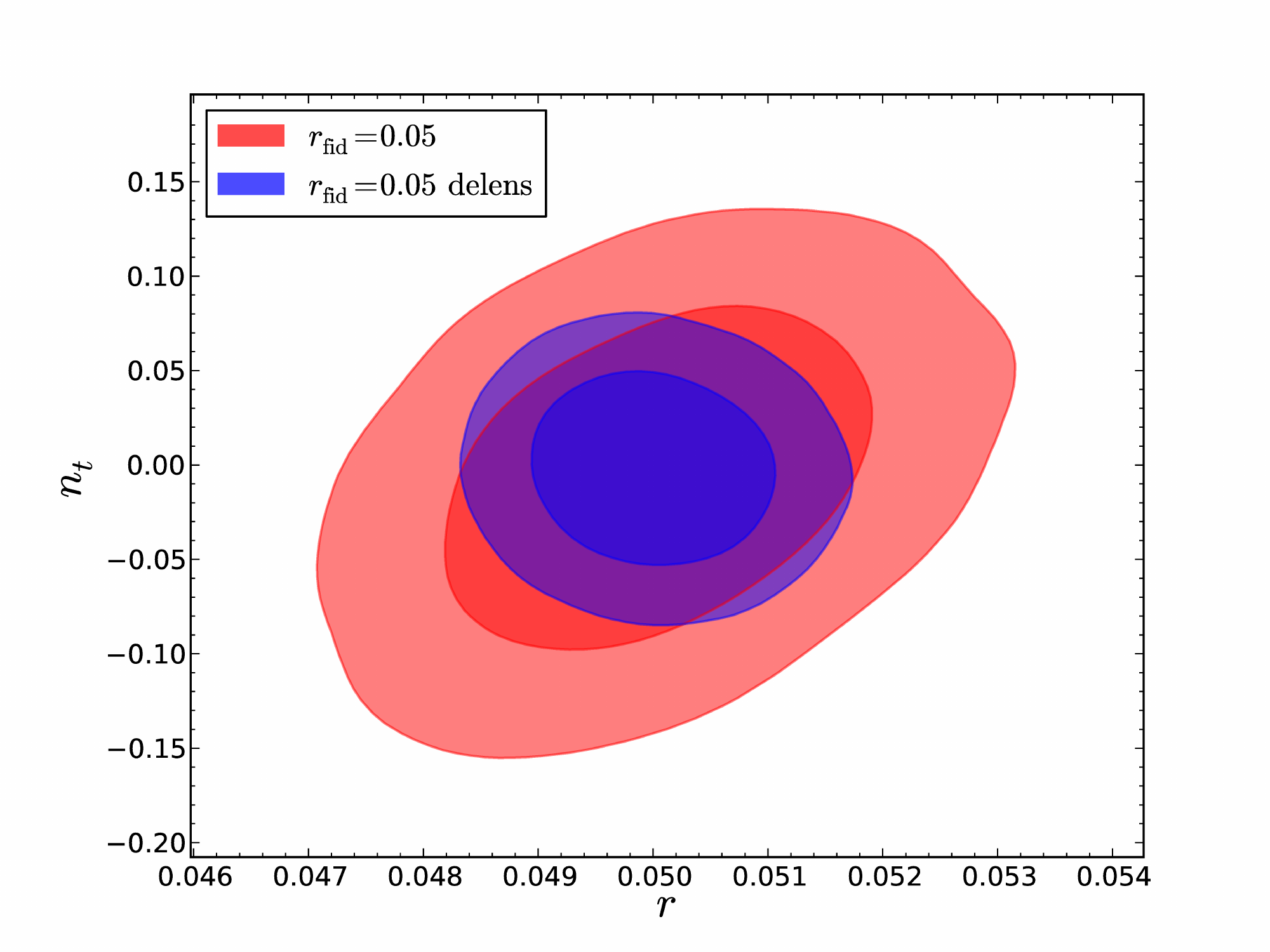}
&\includegraphics[width=\columnwidth]{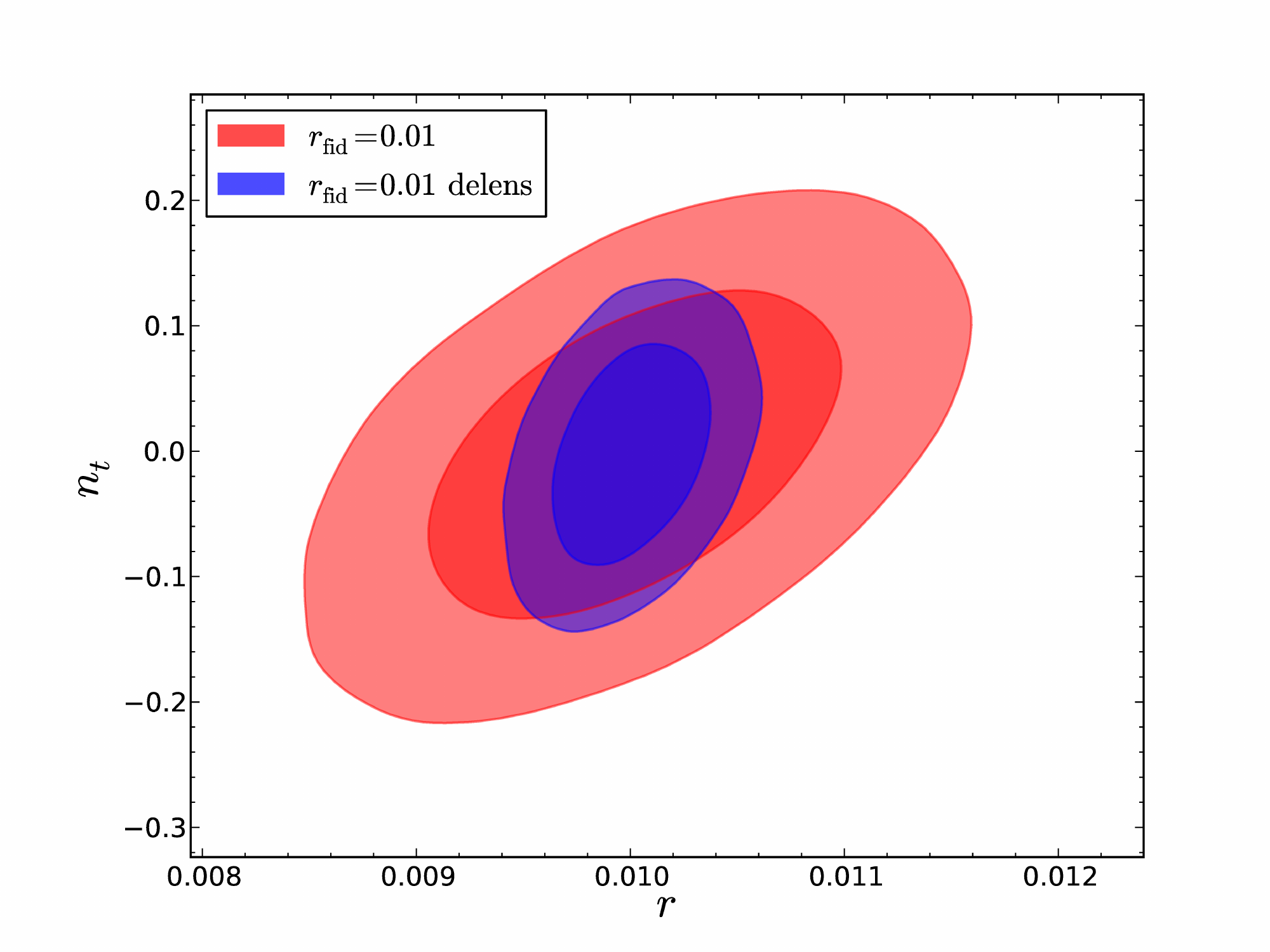}
\end{tabular}
\end{center}
\caption{\footnotesize{Forecasts for $r$ and $\nt$ with two different fiducials: $r = 0.05$ (left panel) and $r = 0.01$ (right panel). In both cases the inflationary consistency relation $\nt = -r/8$ has been assumed. The corresponding $\limit{95}$ 
limits for $r$ and $\nt$ are reported in \tab{forecasts-1}.}}
\label{fig:double_forecast_r_nt}
\end{figure*}

\begin{figure}[!hbtp]
\includegraphics[width=0.48\textwidth]{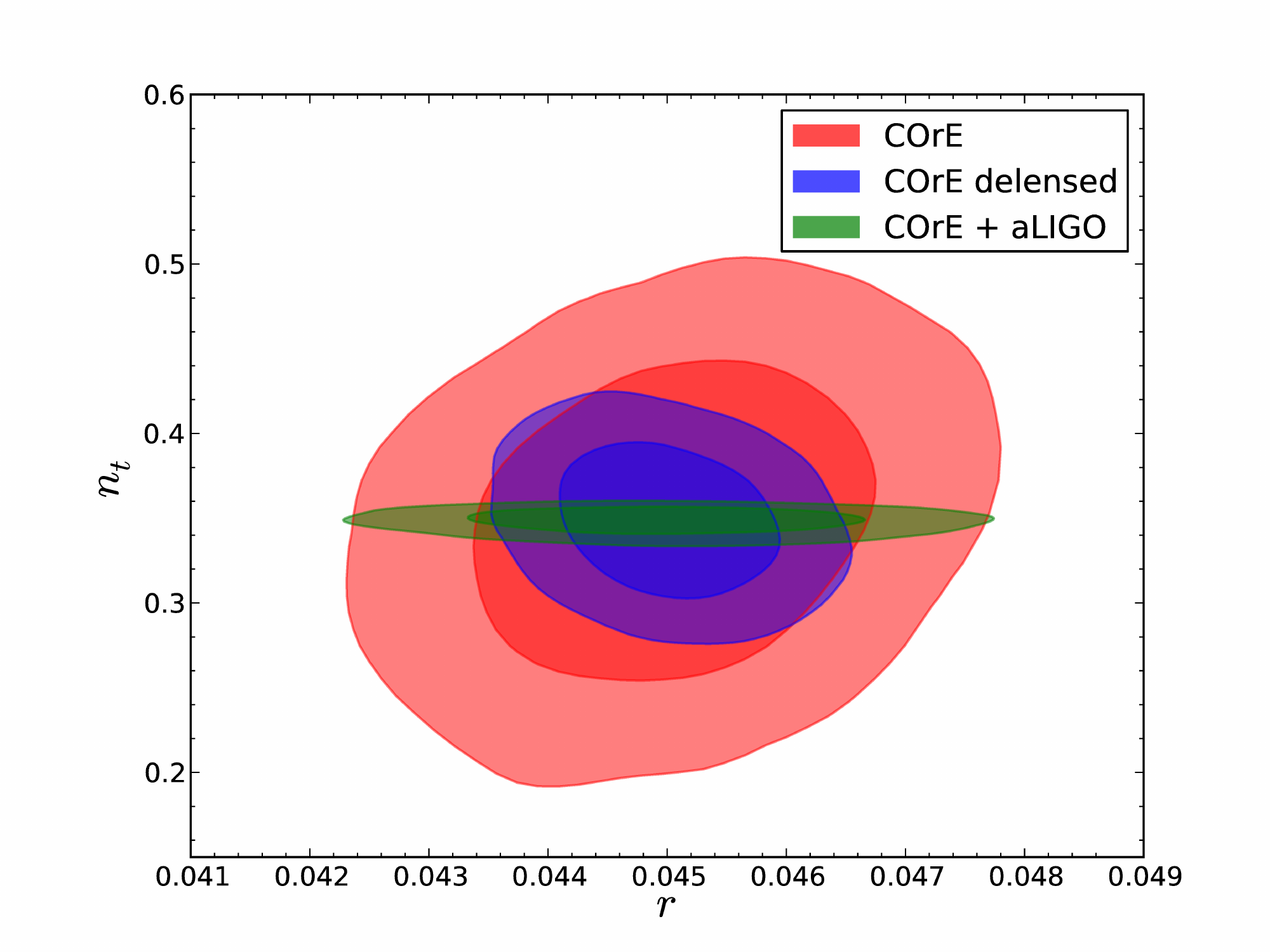}
\caption{\footnotesize{Two-dimensional posterior distributions for $r$ and $\nt$ from COrE (with and without delensing) and COrE + AdvLIGO. The fiducial values are fixed to the best-fit of the \planckBKPLIGO~analysis (\ie $r = 0.045$ and $\nt = 0.35$), and the $\limit{95}$ constraints are reported in \tab{forecasts-2}. We preferred to not include the contours from \planckBKP~+~AdvLIGO (which extend outside of the frame of this plot) to better show the improvement from COrE to COrE + AdvLIGO.}}
\label{fig:forecast_r045_nt35_r_nt}
\end{figure}

\noindent The results of our forecasts are reported in Figs.~\ref{fig:double_forecast_r_nt}, \ref{fig:forecast_r045_nt35_r_nt} and Tabs.~\ref{tab:forecasts-1}, \ref{tab:forecasts-2}.
For our first forecast we assume no detection of $\Omega_\mathrm{GW}$ in the future interferometer experiment AdvLIGO. The datasets we consider are the combination of current CMB measurements (\planckBKP~and \emph{Planck} + BK14) and AdvLIGO experiment. Comparing the results obtained with the current data alone and in combination with AdvLIGO (Tabs.~\ref{tab:results-current}, \ref{tab:results-current-BK}), we see that the constraining power of the next generation of direct detection experiments will be similar to what can be obtained by CMB experiments alone when the contribution $\ngw$ to $\neff$ is included.

We have then considered two fiducial cosmologies, one with $r = 0.05$ and one with $r = 0.01$: in both cases we have taken a fiducial value of the tilt given by $r = -\nt/8$. 
The 
two-dimensional posteriors in the $r$ - $\nt$ plane (\fig{double_forecast_r_nt}) confirm what has been said in \sect{method-delensing}: 
when lensing $B$-modes are removed, one is able to disentangle the effects of $r$ and $\nt$ on the tensor $B$-mode spectrum (since more scales become available and one can distinguish a tilted spectrum from one that is simply rescaled by $r$).

\tab{forecasts-1} shows that COrE will be able to measure $r=0.01$ with a relative uncertainty of order 
\num{3d-2} (\num{d-2} with $10\%$ delensing). On the other hand 
it also shows that, even when delensing is considered, \textcolor{black}{COrE will not be able to probe the inflationary consistency relation with high enough accuracy to pin down single-field slow-roll inflation as the mechanism for the generation of primordial perturbations: in fact we see that $\sigma_{\nt}/\nt$ will be very large, of order \num{10} for the $r = 0.05$ fiducial, and of order \num{100} for the $r = 0.01$ one}. This tells us that the range of scales probed by the CMB will not be sufficient to test 
the scale dependence of the tensor spectrum in the next future: combining CMB measurements with direct detection experiments will be necessary.

Finally, we assume the best-fit from the \planckBKPLIGO~dataset, \ie $r = 0.045$, $\nt = 0.35$, as our fiducial model 
\cite{footnoteBK14A}: the simulated datasets used are \planckBKP~+AdvLIGO, COrE, COrE with $10\%$ delensing and COrE + AdvLIGO (without delensing). We have chosen a ground-based direct detection experiment as additional observable because the lever arm with the scales probed by CMB anisotropies is the strongest available (see \fig{scales}). Besides, since AdvLIGO will put constraints directly on $\Omega_\tu{GW}$, it will be less dependent on the underlying cosmological models than observables like $\mu$-distortions.

The results are reported in \fig{forecast_r045_nt35_r_nt} and \tab{forecasts-2}. The first thing to notice is that \planckBKP~+~AdvLIGO will give a detection of the tensor tilt, while at $\limit{95}$ we will still have only upper bounds on $r$. This is due to the fact that AdvLIGO will actually be able to detect the stochastic background of GWs for these fiducial values of the tensor parameters: however, the tensor power at CMB scales is still too low for \planckBKP~to have a detection of $r$.

Comparing the forecasts from COrE 
\cite{footnoteCMBjust} with those from the combination \planckBKP~+~AdvLIGO, we see that \textcolor{black}{COrE + $10\%$ delensing will 
result in better constraints on the 
tensor parameters than what can be obtained from the evolution of LIGO to AdvLIGO.} 
This is worthy of notice also because the constraints from COrE, 
that will be derived using data from a single experiment (and then with better control of systematics), will be more reliable. 

We also see that combining COrE with the improved version of LIGO will allow to obtain tighter constraints on the tensor tilt than those coming from to COrE alone, even if a $10\%$ delensing is taken into account. More precisely, there is roughly a factor of \num{5} improvement of $\sigma_{\nt}$. On the other hand, we see that the bounds on the tensor-to-scalar ratio are basically unaffected if we add AdvLIGO to the forecast.

We conclude this section with a brief discussion about inflationary models: more precisely about the possibility of having a model 
with $r$ and $\nt$ equal to the best-fit from the \planckBKPLIGO~dataset. One of the main features of single-field slow-roll 
is the presence of the so-called consistency relation: 
it relates the tilt of the primordial tensor spectrum 
to the Hubble slow-roll parameter during inflation, $\epsilon_H\equiv-\dot{H}/H^2$, by 
\begin{equation}
\label{eq:consistency}
\nt = -2\epsilon_H\,\,.
\end{equation}
Single-field slow-roll models do not violate the Null Energy Condition (NEC) 
$\dot{H} < 0$, thus predicting a 
red tensor spectral index $\nt < 0$. 
It is possible, however, to construct models that violate the NEC and lead to a blue 
$\nt$ without incurring in instabilities, like G-inflation \cite{Kobayashi:2010cm} and Ghost inflation \cite{Creminelli:2006xe}. While in Ghost inflation gravitational waves are predicted to be completely unobservable, G-inflation can 
give $r = \mathcal{O}(\num{d-2})$, $\nt = \mathcal{O}(\num{d-1})$: however, it also predicts that the scalar and tensor modes tilt towards the same direction \cite{Wang:2014kqa}, and a blue $\ns>1$ is well excluded by current data (see \tab{fiducials}). 

If these inflationary models are hard-pressed to accomodate such values of the tensor tilt and the tensor-to-scalar ratio, there are other scenarios that can predict a blue $\nt$ while keeping the scalar sector in accord with observations:
\begin{itemize}[leftmargin=*]
\item particle or string sources produced during inflation 
can generate blue tensor modes consistent with the constraints from scalar fluctuations \cite{Senatore:2011sp, Wang:2014kqa}; 
\item gauge field production in axion inflation \cite{Barnaby:2011qe, Mukohyama:2014gba} 
can also lead to blue tensor spectra;
\item it is possible to violate the tensor consistency relation 
also with higher-curvature corrections to the gravitational effective action (coming, \emph{e.g.}, from string theory) \cite{Kaloper:2002uj, Baumann:2015xxa}. In these cases, a time-dependent speed of sound of tensor perturbations changes the consistency relation to
\begin{equation}
\label{eq:DanielConsRel}
\nt = -2\epsilon_H + \mathcal{B}\sqrt{\epsilon_H}\,\,.
\end{equation}
In \cite{Baumann:2015xxa} the authors show that it is possible to make the factor $\mathcal{B}$ positive and 
of order one, and therefore have 
$\nt\lesssim\mathcal{O}(\num{d-1})$.
\end{itemize}
We refer to to \sect{appendix-blue} for a more detailed discussion about these models.

\begin{table}[!hbtp]
\begin{center}
\begin{tabular}{lcc}
\toprule
\horsp
\vertsp  $r$ \vertsp $\nt$ \\
\hline
\horsp
fiducial   \vertsp  $0.05$ \vertsp $-r/8 = -0.00625$ \\
\horsp
COrE  \vertsp $0.0500\pm0.0012$ \vertsp $ -0.0072_{-0.1143}^{+0.1108}$ \\
\horsp
COrE, delens.  \vertsp $0.05000\pm0.00066$ \vertsp $-0.0023 _{-0.0640}^{+0.0632}$ \\
\hline
\horsp
fiducial  \vertsp $0.01$ \vertsp $-r/8 = -0.00125$ \\
\horsp
COrE  \vertsp $0.01001\pm0.00061$ \vertsp $-0.0024 _{-0.1637}^{+0.1597}$ \\
\horsp
COrE, delens.   \vertsp $0.01000\pm0.00024$ \vertsp $-0.0019 _{-0.1088}^{+0.1074}$ \\
\botrule
\end{tabular}
\caption{\footnotesize{Future constraints at $\limit{95}$ on the tensor-to-scalar ratio $r$ and the tensor spectral index $\nt$ from a COrE-like mission (with and without $10\%$ delensing). 
In none of this cases the contribution of $\ngw$ to $\neff$ has been included.}}
\label{tab:forecasts-1}
\end{center}
\end{table}

\begin{table}[!hbtp]
\begin{center}
\begin{tabular}{lcc}
\toprule
\horsp
\vertsp  $r$ \vertsp $\nt$ \\
\hline
\horsp
fiducial  \vertsp $ 0.045$ \vertsp $0.35$ \\
\horsp
\emph{P} + BKP~+ aLIGO \vertsp $< 0.095$ \vertsp $0.354\pm0.020$ \\
\horsp
COrE  \vertsp $0.0450\pm0.0011$ \vertsp $0.348 \pm 0.061$ \\
\horsp
COrE, delens.   \vertsp $0.04500 \pm 0.00060$ \vertsp $0.350 \pm 0.029$ \\
\horsp
COrE + aLIGO   \vertsp $0.0450 \pm 0.0010$ \vertsp $ 0.3483 \pm 0.0053$ \\
\botrule
\end{tabular}
\caption{\footnotesize{
$\limit{95}$ constraints on $r$ and $\nt$ 
from \planckBKP~+ AdvLIGO (denoted by \emph{P} + BKP~+ aLIGO), COrE alone (with and without delensing), and from COrE + AdvLIGO, for a fiducial equal to the best-fit of the \planckBKPLIGO~analysis of \sect{results-current} (\ie $r = 0.045$, $\nt = 0.35$).}}
\label{tab:forecasts-2}
\end{center}
\end{table}

\section{Conclusions}
\label{sec:conclusions}

\noindent In this paper we investigate the constraints on the primordial tensor power spectrum, assuming that it is described by a power law with tilt $\nt$ and tensor-to-scalar ratio $r$, normalized at a pivot scale $k = 0.01\,\mpc^{-1}$. We 
compare the bounds from Cosmic Microwave Background temperature and polarization anisotropies alone with those obtained by adding to the analysis CMB spectral distortions (FIRAS), pulsar timing (European Pulsar Timing Array), and direct detection experiments such as LIGO-Virgo: we find that the gradually stronger lever arm allows to increase the sensitivity from $r < 0.089$, $\nt = 1.7_{-2.0}^{+2.2}$ (\planckBKP, $\limit{95}$) to $r < 0.085$, $\nt = 0.04_{-0.85}^{+0.61}$ (\planckBKPLIGO, $\limit{95}$).

Taking into account the contribution of gravitational waves 
to the effective number of relativistic degrees of freedom $\neff = 3.046 + \ngw$, and the subsequent
effect of an increased radiation energy density on  
CMB angular spectra and 
primordial abundances, the bounds on $r$ and $\nt$ further improve, arriving at $r < 0.081$, $\nt = -0.05_{-0.84}^{+0.58}$ (\planckBKP, $\limit{95}$). These limits on the tensor parameters are stronger than what results from the \planckBKPLIGO~dataset: it must be kept in mind, however, that one must make an explicit assumption about the scale ($k_\tu{UV}$) beyond which primordial GWs are not produced, in order to express $\ngw$ in terms of the primordial tensor spectrum. 
Even by choosing this cutoff to be the scale crossing the horizon at the end of inflation, there is still freedom to vary it due to the uncertainties on the reheating mechanism. In this paper we make the assumption $k_\tu{UV} = \num{d23}\,\mpc^{-1}$, which corresponds to having an instantaneous transition to radiation dominance after inflation ends: this is a conservative choice with respect to other works (which choose, {\it e.g.}, $k_\tu{UV}$ 
equal to the inverse Planck length), but can lead to an overestimation of $\ngw$ in the case of non-instantaneous reheating.

\tblack{If the recently released $95\,\mathrm{GHz}$ data from \emph{Keck Array} are added to the analysis, we find that, while the constraints on the tilt do not get appreciably better, the bounds on the tensor-to-scalar ratio are improved up to $\sim 20\%$: 
for the \emph{Planck} + BK14 + LIGO-Virgo dataset we find $r < 0.067$ ($\limit{95}$), while the \emph{Planck} + BK14 + EXT dataset gives $r < 0.061$ ($\limit{95}$).}

We show that, even with 
the $10\times$ improvement in sensitivity of the upcoming AdvLIGO, the constraints from CMB anisotropies alone will still be stronger than those coming from 
interferometers, if the contribution $\ngw$ to $\neff$ is 
considered.

In the absence of detection, 
the posterior probability distribution for the tensor tilt will 
depend strongly on the prior: for this reason we have also investigated how a future COrE-like CMB mission will be able to constrain $r$ and $\nt$, in the case of fiducial cosmologies with $r$ of order \num{d-2} and $\nt = -r/8$ (as per 
inflationary consistency relation). This value of $r$ has been chosen because it is high enough to be in the reach of not-so-distant experiments like AdvACT.

We find that COrE will measure $r$ with a $\sigma_r/r$ of order \num{d-2}, while the relative uncertainty on $\nt$ will be much larger (of order \num{10} for the $r = 0.05$ fiducial and of order \num{100} for the $r = 0.01$ one). Subtracting lensing $B$-modes to 10\% of their power (an feasible goal for experiments 
where noise can be brought down to $\sim 1\,\mu\K\cdot\mathrm{arcmin}$ after component separation) does little to improve these constraints: however, delensing \textcolor{black}{allows a ``CMB-only'' mission to become competitive with 
the combination of \emph{Planck} + BKP (BK14)~and the upgrade of LIGO, \ie AdvLIGO.}

Finally, we consider a fiducial model where $r$ and $\nt$ are those preferred by the \planckBKPLIGO~dataset. We compare the forecasts for COrE 
with those for COrE combined with AdvLIGO, and we find that adding AdvLIGO will result in a order \num{5} improvement on $\sigma_{\nt}$ with respect to constraints from CMB anisotropies alone, even if we assume that a $10\%$ delensing will be carried out.

We conclude the summary of our results briefly discussing the forecasts for PIXIE \cite{Kogut:2011xw}. PIXIE will provide both photometry and spectrometry: therefore it will allow to combine 
temperature and polarization constraints with those from spectral distortions, while reducing the risk of 
systematics 
caused by the combination of datasets from different experiments. However, we have found that adding the  
bounds on $\mu$-distortions will not 
lead to a decisive improvement: the reason is that $\braket{\mu}$ is dominated by the contribution of scalar perturbations for the values of $r$ and $\nt$ allowed by the constraints from photometry alone. So, since the main constraining power will  
come from temperature and polarization 
anisotropies, we preferred to focus on 
COrE,  whose primary goal will be to improve sensitivity to these two observables.

An interesting development of this work would be to consider upcoming experiments that will measure the expansion history more accurately, and reduce the error on the number $\neff$ of massless degrees of freedom (see for example \cite{Benson:2014qhw}: Fig.~2 -- right panel). In fact, a non-detection of extra radiation would 
impose strong constraints on the contribution of GWs to $\neff$, and then to the tensor parameters $r$ and $\nt$ \cite{footnote7}. 

Another possible development would be to consider a different parametrization of the primordial tensor power spectrum. The assumption of having a power law spectrum down to the scales probed by direct detection experiments is very strong: in the inflationary theory, the more one approaches the end of inflation, the more power spectra will the deviate by the slow-roll result \cite{footnoteENRICO}. Therefore, while writing $\Delta^2_\mathrm{t}(k)$ as a power law is still acceptable when one considers a small range of scales (like the one probed by the CMB), the inclusion of observables which probe scales up to $k\approx\num{d19}\,\mpc^{-1}$ can become at odds with this simple parametrization. One possibility to avoid this problem would be to write $\Delta_\tu{t}^2(k)$ as a step function, and take its $k$-bins to be those shown in \fig{scales}.

While we were completing this work, \cite{Remazeilles:2015hpa} and \cite{Lasky:2015lej} appeared to the arXiv. In the first paper, the authors study how 
the search for the primordial $B$-mode signal will be affected by any imperfect modeling of foregrounds, focusing on their impact on the tensor-to-scalar ratio. In this work we have assumed that foregrounds can be characterized well enough to be taken below 
instrumental noise: we will investigate the effect of foreground modeling for forecasts on the tensor tilt in a future work. In the second paper, the authors also combine bounds on the tensor tilt coming from CMB measurements (direct and indirect, through the effect of GWs on $\neff$), together with pulsar timing and direct detection experiments: we find that, when direct comparison is possible, the results of our works overlap. We also note that \cite{Lasky:2015lej} takes an alternative approach regarding the prior-dependence of the bounds in the absence of detection: they choose a logarithmic prior on the tensor-to-scalar ratio.

\section*{Acknowledgments}

\noindent We would like to thank 
Daniel Baumann, Josquin Errard, Rishi Khatri, Massimiliano Lattanzi and Enrico Pajer 
for useful discussions and comments on the draft.  We would like to thank Antony Lewis for the use of the numerical codes \texttt{cosmomc} and \texttt{camb}. We acknowledge support by the research grant Theoretical Astroparticle Physics number 2012CPPYP7 under the program PRIN 2012 funded by MIUR and by TASP, iniziativa specifica INFN.

\section{Appendix}
\label{sec:appendix}

\subsection{Horizon size at the end of inflation}
\label{sec:appendix-reheating}

\noindent One starts from the following identity for the number $N_\star$ of e-folds of inflation after a scale $k_\star$ has left the horizon (see \cite{Adshead:2010mc} for a derivation)
\begin{equation}
\label{eq:app-1}
\begin{split}
N_\star&\equiv\log\frac{a_\tu{end}}{a_\star} \\
&= -\log\frac{k_\star}{H_0} + \log\frac{H_\star}{H_0} + \log\frac{a_\tu{end}}{a_\tu{reh}} + \log\frac{a_\tu{reh}}{a_0}\,\,,
\end{split}
\end{equation}
where $t_\tu{reh}$ marks the transition to radiation dominance. Then, by taking $k_\star = k_\tu{end}$ (\ie $N_\star = 0$) and making the standard assumption of reheating being a period of matter domination, one can show that
\begin{equation}
\label{eq:app-2}
\begin{split}
\frac{k_\tu{end}}{\mpc^{-1}} = T_\tu{CMB}\exp\bigg[&\log\sqrt[3]{\beta} - \log\sqrt{3} + \log\sqrt[3]{\alpha^2} \\
&+ \log\sqrt[3]{\frac{\pi^2}{45}g_*(T_\tu{CMB})}\bigg]\,\,,
\end{split}
\end{equation}
where $E_\tu{end} = (\alpha\mpl)^4$ and $T_\tu{reh} = \beta\mpl$ are the energy density at the end of inflation and the temperature of the universe at the beginning of radiation dominance, respectively. 

Now: plugging in numerical values for the CMB temperature at the present time ($T_\tu{CMB}\approx\num{2.7}\,\K$), 
and assuming to have instant reheating ($\alpha = \beta$) at the GUT scale $E_\tu{end}\approx\num{d16}\,\gev$ ($\alpha\approx\num{d-2}$), the result is $k_\tu{end}\approx\num{2d23}\,\mpc^{-1}$ as reported in the main text.

\subsection{Forecasting method}
\label{sec:appendix-forecasting}

\noindent To describe the forecasting method used throughout this paper, we focus for simplicity on 
a single anisotropy spectrum of the Cosmic Microwave Background: the generalization to the full $T$, $E$, $B$ spectra is straightforward \cite{Perotto:2006rj}. We start from the expression for the likelihood of a full-sky experiment, \ie (disregarding factors of $2\pi$) 
\begin{equation}
\label{eq:24}
\mathcal{L} = \frac{1}{\sqrt{\det(C_S + C_N)}}e^{-\frac{1}{2} \Delta^T\cdot (C_S + C_N)^{-1}\cdot\Delta}\,\,,
\end{equation}
with data points $\Delta^i_{\ell m} = s^i_{\ell m} + n^i_{\ell m}$ at each pixel (labelled by $\ell m$ since we work in  
harmonic space) and each frequency channel $i$. The signal $s^i_{\ell m}$ is given by $\hat{W}^i_c \hat{a}^c_{\ell m}$, where 
the $\hat{a}_{\ell m}$ are the harmonic  
coefficients of the (beam-smoothed) temperature anisotropy for each component $c$ (\ie CMB + foregrounds such as dust, synchrotron, etc.), and the shape vector $\hat{W}_c^i$ provides the frequency dependence of each component (note that in writing \eq{24} we assume that each component is Gaussian). 
In these formulas we have used a ``hat'' symbol to denote that the $W_c$ and $a^c_{\ell m}$ are those of the specific realization we observe, following \cite{Bond:1998qg}. 

We will assume isotropic white noise in each channel, \ie
\begin{equation}
\label{eq:25}
\braket{n^i_{\ell m}(n^j_{\ell' m'})^*} = w^{-1}_{(i)}\delta^{ij}\delta_{\ell\ell'}\delta_{mm'}\,\,.
\end{equation}
Assuming statistical isotropy, 
the signal covariance matrix will be block diagonal 
in harmonic space: therefore the expression for the log-likelihood $L\equiv -2\log\mathcal{L}$ 
becomes
\begin{equation}
\label{eq:26}
\begin{split}
L = \sum_\ell (2\ell+1)&\bigg\{\Tr\bigg[\frac{\sum_{m = -\ell}^\ell \Delta^i_{\ell m} (\Delta^j_{\ell m})^*}{W^i_c C_{\ell}^{cc'}W^j_{c'} + N_\ell^{ij}}\bigg] \\
&+ \log\det\bigg[W^i_c C_{\ell}^{cc'}W^j_{c'} + N_\ell^{ij}\bigg]\bigg\}\,\,.
\end{split}
\end{equation}
In this equation we denote by $\Tr$ the trace over the frequency channels, and all terms with $i,j$ indices are understood as matrices. Besides we have defined the noise bias as
\begin{equation}
\label{eq:27}
N_\ell^{ij}\equiv N^{(i)}_\ell\delta^{ij} = w^{-1}_{(i)}e^{\sigma_{(i)}^2\ell(\ell+1)}\delta^{ij}\,\,,
\end{equation}
for a gaussian beam of beam-size variance $\sigma_{(i)}$. 
\eq{26} is the expression of the CMB likelihood, once we are given a map $\Delta^i_{\ell m}$. In our forecasts, however, we do not construct explicitly a map, but we use the 
estimator that would be made from such a map, \ie
\begin{equation}
\label{eq:28}
\hat{D}_\ell^{ij} = \sum_{m = -\ell}^\ell \Delta^i_{\ell m} (\Delta^j_{\ell m})^*\equiv \hat{W}^i_c\hat{C}^{cc'}_\ell \hat{W}_{c'}^j + N_\ell^{ij}\,\,,
\end{equation}
where the hats 
denote the fact that the cosmological parameters are fixed to their ``true'' values. Therefore our expression for $L$ becomes
\begin{equation}
\label{eq:29}
\begin{split}
L = \sum_\ell (2\ell+1)&\bigg\{\Tr\bigg[\frac{\hat{W}^i_c \hat{C}_{\ell}^{cc'}\hat{W}^j_{c'} + N_\ell^{ij}}{W^i_c C_{\ell}^{cc'}W^j_{c'} + N_\ell^{ij}}\bigg] \\
&+ \log\det\bigg[W^i_c C_{\ell}^{cc'}W^j_{c'} + N_\ell^{ij}\bigg]\bigg\}\,\,.
\end{split}
\end{equation}
Given fiducial cosmological parameters (which we assume are the ones describing the true universe) and the beam and noise specifications of the experiment (that are given in \tab{COrE-specifics} for a COrE-like experiment), one can construct 
the likelihood for CMB anisotropies, 
and then use it in a Monte Carlo Markov Chain exploration of parameter space. 

\eq{29} simplifies a little if we can consider the case of only one component (the CMB) and forget about foregrounds: however, one has still to take into account both auto- and cross-channel power spectra. For $N_c$ channels with the same noise level, considering both auto and cross power spectra is equivalent to 
have one 
frequency channel with an effective noise power spectrum lower by a factor $N_c$. One can generalize these considerations to the case of channels with different noise levels. The optimal channel combination results in having an 
effective noise bias $N_\ell$ given by \cite{Bond:1997wr, Efstathiou:1998xx, Verde:2005ff}
\begin{equation}
\label{eq:30}
\begin{split}
N_\ell &= \bigg(\sum_i \frac{1}{N^{(i)}_\ell}\bigg)^{-1} \\
&= \bigg(\sum_i w_{(i)}e^{-\sigma_{(i)}^2\ell(\ell+1)}\bigg)^{-1}\,\,.
\end{split}
\end{equation}

In reality the presence of foregrounds limits our ability  
of extracting the CMB signal from the data, and a full likelihood analysis should take them into account. Fortunately, each component scales differently in frequency (\ie, every foreground has a different shape $W_c$): therefore it is possible to separate them using maps at different frequencies \cite{Hobson:1998td, Baccigalupi:2002bh}. 
This foreground subtraction will be the source of additional noise, 
depending on the level of foreground removal, which 
will contribute to the noise bias $N_\ell$ (we refer to \cite{Verde:2005ff} for a more detailed analysis). In our forecasts we consider the case where this additional noise 
is much smaller 
than the instrumental noise 
of \eq{27}, so that \eq{30} is recovered.

Therefore, normalizing the likelihood such that $L = 0$ at the fiducial values of the cosmological parameters, we have that (for $BB$ spectra only) 
\begin{equation}
\label{eq:31}
\begin{split}
L = \sum_\ell (2\ell+1)&\bigg[-1 + \frac{\hat{C}^\tu{tens}_\ell + \hat{C}^\tu{lens}_\ell + N_\ell}{C^\tu{tens}_\ell + C^\tu{lens}_\ell + N_\ell} \\
&+ \log\bigg(\frac{C^\tu{tens}_\ell + C^\tu{lens}_\ell + N_\ell}{\hat{C}^\tu{tens}_\ell + \hat{C}^\tu{lens}_\ell + N_\ell}\bigg)\bigg]\,\,,
\end{split}
\end{equation}
where the ``tens'' superscript labels the $BB$ spectrum from primordial tensor modes, and the ``lens'' superscript labels $B$-modes due to gravitational lensing. As explained in \sect{method-delensing}
, delensing will be implemented by rescaling $C^\tu{lens}_\ell$ and $\hat{C}^\tu{lens}_\ell$ as
\begin{subequations}
\label{eq:31-bis}
\begin{align}
&C^\tu{lens}_\ell\to\num{0.1}\times C^\tu{lens}_\ell\,\,, \label{eq:31-bis-1} \\
&\hat{C}^\tu{lens}_\ell\to\num{0.1}\times\hat{C}^\tu{lens}_\ell\,\,. \label{eq:31-bis-2}
\end{align}
\end{subequations}

We will use the likelihood of \eq{31} for the MCMC exploration of parameter space, with one additional caveat: in the case of a non full-sky experiment (where only part of the sky is observed or can be used for cosmology) 
not all modes are available for the analysis, and \eq{29} does not hold. One can capture this effect by introducing the $\fsky$ parameter, which reduces the available modes by
\begin{equation}
\label{eq:32}
\sum_\ell (2\ell+1)\to\sum_\ell (2\ell+1)\times\fsky\,\,.
\end{equation}
We 
follow this approach in our forecast method (with $\fsky = \num{0.8}$ for the COrE satellite), and we refer the reader to \cite{Wandelt:2000av} for better approximations.

\subsection{Inflationary models and blue $n_\mathrm{t}$}
\label{sec:appendix-blue}

\noindent In this appendix we briefly review two models that can produce blue gravity waves, while keeping the scalar sector in accord with observational constraints.

In the model of \cite{Barnaby:2011qe},
the inflaton $\phi$ is an axion, and the breaking of its shift symmetry $\phi\to\phi+c$ allows for a coupling with a gauge field $A_\mu$ (more precisely with its field strength $F_{\mu\nu}$) given by
\begin{equation}
\label{eq:app3-1}
\mathcal{L}\supset -\frac{\alpha}{4f}\phi F_{\mu\nu}\tilde{F}^{\mu\nu}\,\,,
\end{equation}
where the dual $\tilde{F}_{\mu\nu}$ is $\epsilon_{\mu\nu\rho\sigma} F^{\rho\sigma}$ and $f$ is the axion decay constant. The parameter controlling the strength of gauge field production is
\begin{equation}
\label{eq:app3-2}
\xi\equiv\frac{\alpha\dot{\braket{\phi}}}{2fH}\,\,,
\end{equation}
and one can show that the leading contribution to the $A_\mu$ occupation numbers goes like $e^{2\pi\xi}/\sqrt{\xi}$. 
The produced gauge fields will act as a source for tensor modes, and the resulting 
tensor power spectrum (for the left and right polarizations $L$ and $R$) will be \cite{Barnaby:2010vf, Sorbo:2011rz, Barnaby:2011vw}
\begin{equation}
\label{eq:app3-3}
\begin{split}
P_\tu{t}^{L,R}(k) &= \frac{H^2}{\pi^2M^2_\tu{P}}\bigg(\frac{k}{k_\star}\bigg)^{\nt}\times \\
&\;\;\;\;\bigg(1 + 2\frac{H^2}{M^2_\tu{P}}f^{L,R}(\xi) e^{4\pi\xi}\bigg)\,\,, 
\end{split}
\end{equation}
where the function $f_L$ ($f_R$) will be of order $\num{4d-7}/\xi^6$ ($\num{9d-10}/\xi^6$) at large $\xi$ \cite{Barnaby:2011vw, footnote-app3-1}. We see from \eq{app3-3} that, in the limit of $\dot{\xi} = 0$, the additional contribution to the tensor power will not give a tilt different from the usual single-field slow-roll result: its only effect will be to enhance the value of $r$ \cite{Barnaby:2011qe}. 
However $\xi$ is increasing during inflation, and one can show that, in the limit where $\delta_\xi\equiv\dot{\xi}/H\xi$ (the fractional variation per Hubble time of $\xi$) is small, the tensor tilt will receive a correction $\approx (4\pi\xi - 6)\delta_\xi$ (see also \cite{Mukohyama:2014gba}). Therefore, for modestly large values of $\xi$, there is room to have $0\lesssim\nt\lesssim\mathcal{O}(1)􏶯$.

The model described in \cite{Baumann:2015xxa}, instead, considers a coupling of the inflaton to the square of the Weyl tensor, \ie
\begin{equation}
\label{app3-4}
\mathcal{L}\supset f(\phi)\frac{W^2}{M^2}\,\,,
\end{equation}
on top of the slow-roll action. Working out the action for 
tensor perturbations shows that they have a non-trivial speed 
sound 
$c_\mathrm{t}$. 
While $c_\mathrm{t}$ does not deviate too much from unity, 
\cite{Baumann:2015xxa} shows that 
it can still have a sizable time dependence. The fractional change per Hubble time of the tensor speed of sound will give then a contribution to the tensor tilt $\nt$, proportional to  
$\sqrt{\epsilon_H}$: therefore the consistency relation of \eq{consistency} 
will not hold in this model. The sign and magnitude of the proportionality factor are determined by the derivative of the function $f$ and from the ratio $H^2/M^2$: with suitable choices for the scale of variation of $f$ and for the scale $M$, the overall contribution can lead to blue 
tilt.


\begin{thebibliography}{99}

\bibitem{deBernardis:2000gy} 
  P.~de Bernardis {\it et al.} [Boomerang Collaboration],
  Nature {\bf 404}, 955 (2000).
  
\bibitem{Eisenstein:2005su} 
  D.~J.~Eisenstein {\it et al.} [SDSS Collaboration],
  Astrophys.\ J.\  {\bf 633}, 560 (2005).

\bibitem{Hinshaw:2012aka} 
  G.~Hinshaw {\it et al.} [WMAP Collaboration],
  Astrophys.\ J.\ Suppl.\  {\bf 208}, 19 (2013).
  
\bibitem{Adam:2015rua} 
  R.~Adam {\it et al.} [Planck Collaboration],
  arXiv:1502.01582 [astro-ph.CO].
  
\bibitem{Guth:1980zm} 
  A.~H.~Guth,
  Phys.\ Rev.\ D {\bf 23}, 347 (1981).
  
\bibitem{Linde:1981mu} 
  A.~D.~Linde,
  Phys.\ Lett.\ B {\bf 108}, 389 (1982).
  
\bibitem{Albrecht:1982wi} 
  A.~Albrecht and P.~J.~Steinhardt,
  Phys.\ Rev.\ Lett.\  {\bf 48}, 1220 (1982).

\bibitem{Baumann:2009ds} 
  D.~Baumann,
  arXiv:0907.5424 [hep-th].

\bibitem{kamion}
  M.~Kamionkowski and E.~D.~Kovetz,
  arXiv:1510.06042 [astro-ph.CO].

\bibitem{Aghanim:2015xee} 
  N.~Aghanim {\it et al.} [Planck Collaboration],
  [arXiv:1507.02704 [astro-ph.CO]].
  
\bibitem{Ade:2015tva} 
  P.~A.~R.~Ade {\it et al.}  [BICEP2 and Planck Collaborations],
  Phys.\ Rev.\ Lett.\  {\bf 114}, no. 10, 101301 (2015).
  
\bibitem{Array:2015xqh} 
  K.~Array {\it et al.} [BICEP2 s Collaboration],
  arXiv:1510.09217 [astro-ph.CO].
  
\bibitem{Ade:2015lrj} 
  P.~A.~R.~Ade {\it et al.} [Planck Collaboration],
  arXiv:1502.02114 [astro-ph.CO].

\bibitem{Allen:1996vm} 
  B.~Allen,
  In *Les Houches 1995, Relativistic gravitation and gravitational radiation* 373-417.

\bibitem{Maggiore:1999vm} 
  M.~Maggiore,
  Phys.\ Rept.\  {\bf 331}, 283 (2000).


\bibitem{Cutler:2002me} 
  C.~Cutler and K.~S.~Thorne,
  gr-qc/0204090.

\bibitem{Moore:2014lga}
  C.~J.~Moore, R.~H.~Cole and C.~P.~L.~Berry,
  Class.\ Quant.\ Grav.\  {\bf 32} (2015) 1,  015014
  doi:10.1088/0264-9381/32/1/015014

\bibitem{Abbott:2007kv} 
  B.~P.~Abbott {\it et al.} [LIGO Scientific Collaboration],
  Rept.\ Prog.\ Phys.\  {\bf 72}, 076901 (2009).

\bibitem{Abadie:2011xta} 
  J.~Abadie {\it et al.} [LIGO Collaboration],
  Nature Phys.\  {\bf 7}, 962 (2011).
  
\bibitem{Acernese:2004jn} 
  F.~Acernese {\it et al.} [VIRGO Collaboration],
  Class.\ Quant.\ Grav.\  {\bf 22}, S869 (2005).
  
\bibitem{TheVirgo:2014hva} 
  F.~Acernese {\it et al.} [VIRGO Collaboration],
  Class.\ Quant.\ Grav.\  {\bf 32}, no. 2, 024001 (2015).
  
\bibitem{AmaroSeoane:2012km} 
  P.~Amaro-Seoane {\it et al.},
  GW Notes {\bf 6}, 4 (2013).
  
\bibitem{AmaroSeoane:2012je} 
  P.~Amaro-Seoane {\it et al.},
  Class.\ Quant.\ Grav.\  {\bf 29}, 124016 (2012).
  
\bibitem{Kawamura:2006up} 
  S.~Kawamura {\it et al.},
  Class.\ Quant.\ Grav.\  {\bf 23}, S125 (2006).
  
\bibitem{Kawamura:2011zz} 
  S.~Kawamura {\it et al.},
  Class.\ Quant.\ Grav.\  {\bf 28}, 094011 (2011).
  
\bibitem{Crowder:2005nr} 
  J.~Crowder and N.~J.~Cornish,
  Phys.\ Rev.\ D {\bf 72}, 083005 (2005).
  
\bibitem{Ferdman:2010xq} 
  R.~D.~Ferdman {\it et al.},
  Class.\ Quant.\ Grav.\  {\bf 27}, 084014 (2010).
  
\bibitem{Fixsen:1996nj} 
  D.~J.~Fixsen, E.~S.~Cheng, J.~M.~Gales, J.~C.~Mather, R.~A.~Shafer and E.~L.~Wright,
  Astrophys.\ J.\  {\bf 473}, 576 (1996).

\bibitem{Ota:2014hha} 
  A.~Ota, T.~Takahashi, H.~Tashiro and M.~Yamaguchi,
  JCAP {\bf 1410}, no. 10, 029 (2014).

\bibitem{Chluba:2014qia} 
  J.~Chluba, L.~Dai, D.~Grin, M.~Amin and M.~Kamionkowski,
  Mon.\ Not.\ Roy.\ Astron.\ Soc.\  {\bf 446}, 2871 (2015).

\bibitem{Hu:1992dc} 
  W.~Hu and J.~Silk,
  Phys.\ Rev.\ D {\bf 48}, 485 (1993).
  
\bibitem{Chluba:2012gq} 
  J.~Chluba, R.~Khatri and R.~A.~Sunyaev,
  Mon.\ Not.\ Roy.\ Astron.\ Soc.\  {\bf 425}, 1129 (2012).
  
\bibitem{Pajer:2013oca} 
  E.~Pajer and M.~Zaldarriaga,
  JCAP {\bf 1302}, 036 (2013).

\bibitem{Hu:1994bz} 
  W.~Hu, D.~Scott and J.~Silk,
  Astrophys.\ J.\  {\bf 430}, L5 (1994).
  
\bibitem{Chluba:2012we} 
  J.~Chluba, A.~L.~Erickcek and I.~Ben-Dayan,
  Astrophys.\ J.\  {\bf 758}, 76 (2012).
  
\bibitem{Mangano:2001iu} 
  G.~Mangano, G.~Miele, S.~Pastor and M.~Peloso,
  Phys.\ Lett.\ B {\bf 534}, 8 (2002).
  
\bibitem{Mangano:2005cc} 
  G.~Mangano, G.~Miele, S.~Pastor, T.~Pinto, O.~Pisanti and P.~D.~Serpico,
  Nucl.\ Phys.\ B {\bf 729}, 221 (2005).

\bibitem{Bowen:2001in} 
  R.~Bowen, S.~H.~Hansen, A.~Melchiorri, J.~Silk and R.~Trotta,
  Mon.\ Not.\ Roy.\ Astron.\ Soc.\  {\bf 334}, 760 (2002).

\bibitem{Hou:2011ec} 
  Z.~Hou, R.~Keisler, L.~Knox, M.~Millea and C.~Reichardt,
  Phys.\ Rev.\ D {\bf 87}, 083008 (2013).

\bibitem{Izotov:2014fga} 
  Y.~I.~Izotov, T.~X.~Thuan and N.~G.~Guseva,
  Mon.\ Not.\ Roy.\ Astron.\ Soc.\  {\bf 445}, no. 1, 778 (2014).
  
\bibitem{Mucciarelli:2014abc} 
  A.~Mucciarelli, L.~Lovisi, B.~Lanzoni, and F.~R.~Ferraro,
  Astrophys.\ J.\  {\bf 786}, no. 1, 14 (2014).

\bibitem{Cooke:2013cba}
  R.~Cooke, M.~Pettini, R.~A.~Jorgenson, M.~T.~Murphy and C.~C.~Steidel,
  Astrophys.\ J.\  {\bf 781} (2014) 1,  31.

\bibitem{Cyburt:2001pq} 
  R.~H.~Cyburt, B.~D.~Fields and K.~A.~Olive,
  Astropart.\ Phys.\  {\bf 17}, 87 (2002).

\bibitem{Ichikawa:2007js} 
  K.~Ichikawa, T.~Sekiguchi and T.~Takahashi,
  Phys.\ Rev.\ D {\bf 78}, 043509 (2008).

\bibitem{Damour:2000wa} 
  T.~Damour and A.~Vilenkin,
  Phys.\ Rev.\ Lett.\  {\bf 85}, 3761 (2000).

\bibitem{Olmez:2010bi} 
  S.~Olmez, V.~Mandic and X.~Siemens,
  Phys.\ Rev.\ D {\bf 81}, 104028 (2010).

\bibitem{Ungarelli:2005qb} 
  C.~Ungarelli, P.~Corasaniti, R.~A.~Mercer and A.~Vecchio,
  Class.\ Quant.\ Grav.\  {\bf 22}, S955 (2005).

\bibitem{Smith:2006xf} 
  T.~L.~Smith, H.~V.~Peiris and A.~Cooray,
  Phys.\ Rev.\ D {\bf 73}, 123503 (2006).

\bibitem{Chongchitnan:2006pe} 
  S.~Chongchitnan and G.~Efstathiou,
  Phys.\ Rev.\ D {\bf 73}, 083511 (2006).
  
\bibitem{Smith:2006nka} 
  T.~L.~Smith, E.~Pierpaoli and M.~Kamionkowski,
  Phys.\ Rev.\ Lett.\  {\bf 97}, 021301 (2006).

\bibitem{Boyle:2007zx} 
  L.~A.~Boyle and A.~Buonanno,
  Phys.\ Rev.\ D {\bf 78}, 043531 (2008).

\bibitem{Stewart:2007fu} 
  A.~Stewart and R.~Brandenberger,
  JCAP {\bf 0808}, 012 (2008).

\bibitem{Camerini:2008mj}
  R.~Camerini, R.~Durrer, A.~Melchiorri, A.~Riotto, R.~Durrer, A.~Melchiorri and A.~Riotto,
  Phys.\ Rev.\ D {\bf 77} (2008) 101301
  [arXiv:0802.1442 [astro-ph]].

\bibitem{Sendra:2012wh} 
  I.~Sendra and T.~L.~Smith,
  Phys.\ Rev.\ D {\bf 85}, 123002 (2012).

\bibitem{DiValentino:2011zz}
  E.~Di Valentino, A.~Melchiorri and L.~Pagano,
  Int.\ J.\ Mod.\ Phys.\ D {\bf 20} (2011) 1183.

\bibitem{Gerbino:2014eqa} 
  M.~Gerbino, A.~Marchini, L.~Pagano, L.~Salvati, E.~Di Valentino and A.~Melchiorri,
  Phys.\ Rev.\ D {\bf 90}, no. 4, 047301 (2014).


\bibitem{Henrot-Versille:2014jua} 
  S.~Henrot-Versille {\it et al.},
  Class.\ Quant.\ Grav.\  {\bf 32}, no. 4, 045003 (2015).

\bibitem{Meerburg:2015zua} 
  P.~D.~Meerburg, R.~Hlo\v{z}ek, B.~Hadzhiyska and J.~Meyers,
  Phys.\ Rev.\ D {\bf 91}, no. 10, 103505 (2015).
  
\bibitem{Huang:2015gka} 
  Q.~G.~Huang and S.~Wang,
  JCAP {\bf 1506}, no. 06, 021 (2015).
  
\bibitem{Nakama:2015nea} 
  T.~Nakama and T.~Suyama,
  Phys.\ Rev.\ D {\bf 92}, no. 12, 121304 (2015).
  
\bibitem{Pagano:2015hma} 
  L.~Pagano, L.~Salvati and A.~Melchiorri,
  arXiv:1508.02393 [astro-ph.CO].
  
\bibitem{Huang:2015gca} 
  Q.~G.~Huang, S.~Wang and W.~Zhao,
  JCAP {\bf 1510}, no. 10, 035 (2015).
  
\bibitem{Kogut:2011xw} 
  A.~Kogut {\it et al.},
  JCAP {\bf 1107}, 025 (2011).
  
\bibitem{Matsumura:2013aja} 
  T.~Matsumura {\it et al.},
  Journal of Low Temperature Physics September 2014, Volume 176,
  Issue 5-6, pp 733-740.

\bibitem{Bouchet:2011ck} 
  F.~R.~Bouchet {\it et al.} [COrE Collaboration],
  arXiv:1102.2181 [astro-ph.CO].
  
\bibitem{TheLIGOScientific:2014jea} 
  J.~Aasi {\it et al.} [LIGO Scientific Collaboration],
  Class.\ Quant.\ Grav.\  {\bf 32}, 074001 (2015).

\bibitem{Calabrese:2014gwa}
  E.~Calabrese {\it et al.},
  JCAP {\bf 1408} (2014) 010.
    
\bibitem{Niemack:2010wz} 
  M.~D.~Niemack {\it et al.},
  Proc.\ SPIE Int.\ Soc.\ Opt.\ Eng.\  {\bf 7741}, 77411S (2010).
  
\bibitem{footnote0}
  Besides, if tensor modes are actually detected, the bounds that one will obtain on the tensor tilt will be prior-independent: see \sect{results} for a discussion.
    
\bibitem{Errard:2015cxa} 
  J.~Errard, S.~M.~Feeney, H.~V.~Peiris and A.~H.~Jaffe,
  arXiv:1509.06770 [astro-ph.CO].
  
\bibitem{Lidsey:1995np} 
  J.~E.~Lidsey, A.~R.~Liddle, E.~W.~Kolb, E.~J.~Copeland, T.~Barreiro and M.~Abney,
  Rev.\ Mod.\ Phys.\  {\bf 69}, 373 (1997).

\bibitem{Boyle:2005se} 
  L.~A.~Boyle and P.~J.~Steinhardt,
  Phys.\ Rev.\ D {\bf 77}, 063504 (2008).

\bibitem{Watanabe:2006qe} 
  Y.~Watanabe and E.~Komatsu,
  Phys.\ Rev.\ D {\bf 73}, 123515 (2006).

\bibitem{Pritchard:2004qp} 
  J.~R.~Pritchard and M.~Kamionkowski,
  Annals Phys.\  {\bf 318}, 2 (2005).
  
\bibitem{footnote1}
  This condition is necessary for $\rho_{\rm GW}$ to redshift as radiation (\ie as $a^{-4}$), 
which in turn has been used in deriving Eqs.~\eqref{eq:Ngw-rho-gstar-f} and \eqref{eq:Ngw-final}.
  
 \bibitem{Smith:2008fea}
  T.~L.~Smith,
  ``The gravity of the situation,''
  
\bibitem{Jeong:2014gna} 
  D.~Jeong, J.~Pradler, J.~Chluba and M.~Kamionkowski,
  Phys.\ Rev.\ Lett.\  {\bf 113}, 061301 (2014).
  
\bibitem{Chluba:2011hw} 
  J.~Chluba and R.~A.~Sunyaev,
  Mon.\ Not.\ Roy.\ Astron.\ Soc.\  {\bf 419}, 1294 (2012).

\bibitem{Weinberg:1971mx} 
  S.~Weinberg,
  Astrophys.\ J.\  {\bf 168}, 175 (1971).
  
\bibitem{Kaiser:1983abc} 
  N.~Kaiser,
  Mon.\ Not.\ Roy.\ Astron.\ Soc.\  {\bf 202}, 1169 (1983).

\bibitem{Hu:1997hp} 
  W.~Hu and M.~J.~White,
  Phys.\ Rev.\ D {\bf 56}, 596 (1997).

\bibitem{footnote2}
  This approximation works well in the tight-coupling regime which, again, is a valid hypothesis at the redshifts when $\mu$-distortions are generated.

\bibitem{Weinberg:2003ur} 
  S.~Weinberg,
  Phys.\ Rev.\ D {\bf 69}, 023503 (2004).

\bibitem{Khatri:2011aj} 
  R.~Khatri, R.~A.~Sunyaev and J.~Chluba,
  Astron.\ Astrophys.\  {\bf 540}, A124 (2012).
  
\bibitem{footnote3}
  When one is dealing with small perturbations of the Planck spectrum, 
linear theory can be used: in the linear limit, then, 
it is possible to show that for large enough frequencies (\ie away from the Rayleigh-Jeans tail) a small $\mu$-distortion with negative sign is a good approximation.

\bibitem{Detweiler:1979abc} 
  S.~Detweiler,
  Astrophys.\ J.\  {\bf 234}, 1100 (1979).
 
\bibitem{Pitkin:2011yk} 
  M.~Pitkin, S.~Reid, S.~Rowan and J.~Hough,
  Living Rev.\ Rel.\  {\bf 14}, 5 (2011).
  
\bibitem{Adhikari:2013kya} 
  R.~X.~Adhikari,
  Rev.\ Mod.\ Phys.\  {\bf 86}, 121 (2014).

\bibitem{Lewis:2002ah} 
  A.~Lewis and S.~Bridle,
  Phys.\ Rev.\ D {\bf 66}, 103511 (2002).
  
\bibitem{Lewis:2013hha} 
  A.~Lewis,
  Phys.\ Rev.\ D {\bf 87}, no. 10, 103529 (2013).
  
  
\bibitem{Kawasaki:1999na} 
  M.~Kawasaki, K.~Kohri and N.~Sugiyama,
  Phys.\ Rev.\ Lett.\  {\bf 82}, 4168 (1999).
  
\bibitem{Kawasaki:2000en} 
  M.~Kawasaki, K.~Kohri and N.~Sugiyama,
  Phys.\ Rev.\ D {\bf 62}, 023506 (2000).
  
\bibitem{deSalas:2015glj} 
  P.~F.~de Salas, M.~Lattanzi, G.~Mangano, G.~Miele, S.~Pastor and O.~Pisanti,
  arXiv:1511.00672 [astro-ph.CO].

\bibitem{Planck:2015xua} 
  P.~A.~R.~Ade {\it et al.}  [Planck Collaboration],
  arXiv:1502.01589 [astro-ph.CO].

\bibitem{Beutler:2011hx} 
  F.~Beutler {\it et al.},
  Mon.\ Not.\ Roy.\ Astron.\ Soc.\  {\bf 416}, 3017 (2011).
  
\bibitem{Ross:2014qpa} 
  A.~J.~Ross, L.~Samushia, C.~Howlett, W.~J.~Percival, A.~Burden and M.~Manera,
  Mon.\ Not.\ Roy.\ Astron.\ Soc.\  {\bf 449}, no. 1, 835 (2015).
  
\bibitem{Anderson:2013zyy} 
  L.~Anderson {\it et al.} [BOSS Collaboration],
  Mon.\ Not.\ Roy.\ Astron.\ Soc.\  {\bf 441}, no. 1, 24 (2014).

\bibitem{Salvati:2015wxa} 
  L.~Salvati, L.~Pagano, R.~Consiglio and A.~Melchiorri,
  arXiv:1507.07243 [astro-ph.CO].
  
\bibitem{Aasi:2014zwg} 
  J.~Aasi {\it et al.} [LIGO Scientific and VIRGO Collaborations],
  Phys.\ Rev.\ Lett.\  {\bf 113}, no. 23, 231101 (2014).
  
\bibitem{Lentati:2015qwp} 
  L.~Lentati {\it et al.},
  Mon.\ Not.\ Roy.\ Astron.\ Soc.\  {\bf 453}, 2576 (2015).
  
  
\bibitem{Lewis:2005tp} 
  A.~Lewis,
  Phys.\ Rev.\ D {\bf 71}, 083008 (2005).
 
\bibitem{core++}
  \url{http://www.core-mission.org}.

\bibitem{Zaldarriaga:1998ar}
  M.~Zaldarriaga and U.~Seljak,
  Phys.\ Rev.\ D {\bf 58} (1998) 023003.
  
\bibitem{Smith:2008an}
  K.~M.~Smith {\it et al.},
  AIP Conf.\ Proc.\  {\bf 1141} (2009) 121.
  
\bibitem{Knox:2002pe} 
  L.~Knox and Y.~S.~Song,
  Phys.\ Rev.\ Lett.\  {\bf 89}, 011303 (2002).
  
\bibitem{Kesden:2002ku} 
  M.~Kesden, A.~Cooray and M.~Kamionkowski,
  Phys.\ Rev.\ Lett.\  {\bf 89}, 011304 (2002).
  
\bibitem{Seljak:2003pn} 
  U.~Seljak and C.~M.~Hirata,
  Phys.\ Rev.\ D {\bf 69}, 043005 (2004).
  
\bibitem{Smith:2010gu} 
  K.~M.~Smith, D.~Hanson, M.~LoVerde, C.~M.~Hirata and O.~Zahn,
  JCAP {\bf 1206}, 014 (2012).
 
\bibitem{Creminelli:2015oda} 
  P.~Creminelli, D.~L.~Nacir, M.~Simonovi\'{c}, G.~Trevisan and M.~Zaldarriaga,
  arXiv:1502.01983 [astro-ph.CO].
  
\bibitem{footnote4}
  We also see from \fig{delensing} that this argument does not hold for a tensor-to-scalar ratio too small, since in that case the primordial $B$-mode angular spectrum $C^\tu{tens}_\ell$ 
is swamped by noise too soon, and the sensitivity to $\nt$ is lost even if we reduce the power of lensing $B$-modes.

\bibitem{footnote-prior}
  As shown in \cite{Creminelli:2015oda}, while a value of $r$ of order \num{d-3} will not be probed by future ground-based experiments, future satellite missions will be able to detect it with a $\sigma_r/r$ of order \num{d-1}.
  
\bibitem{footnote5}
  This mechanism is the same that was proposed in \cite{Gerbino:2014eqa} to explain the tension between \planck~data and the claim from the BICEP2 collaboration of $r\approx 0.2$ at $0.05\,\mpc^{-1}$.
  
  
\bibitem{footnoteBK14A}
  Since using the BK14 dataset 
does not change significantly the best-fit 
of the combined analysis with LIGO-Virgo, we will not consider the forecast for \emph{Planck} + BK14 + AdvLIGO.
  
\bibitem{footnoteCMBjust}
  Which, we note, are obtained with a ``just CMB'' dataset, \ie by looking only at the effects of primordial tensor modes on CMB temperature and polarization anisotropies.
  
\bibitem{Kobayashi:2010cm} 
  T.~Kobayashi, M.~Yamaguchi and J.~Yokoyama,
  Phys.\ Rev.\ Lett.\  {\bf 105}, 231302 (2010).
  
\bibitem{Creminelli:2006xe} 
  P.~Creminelli, M.~A.~Luty, A.~Nicolis and L.~Senatore,
  JHEP {\bf 0612}, 080 (2006).
  
\bibitem{Wang:2014kqa} 
  Y.~Wang and W.~Xue,
  JCAP {\bf 1410}, no. 10, 075 (2014).
  
\bibitem{Senatore:2011sp} 
  L.~Senatore, E.~Silverstein and M.~Zaldarriaga,
  JCAP {\bf 1408}, 016 (2014).
  
\bibitem{Barnaby:2011qe} 
  N.~Barnaby, E.~Pajer and M.~Peloso,
  Phys.\ Rev.\ D {\bf 85}, 023525 (2012).
  
\bibitem{Mukohyama:2014gba} 
  S.~Mukohyama, R.~Namba, M.~Peloso and G.~Shiu,
  JCAP {\bf 1408}, 036 (2014).
  
\bibitem{Kaloper:2002uj} 
  N.~Kaloper, M.~Kleban, A.~E.~Lawrence and S.~Shenker,
  Phys.\ Rev.\ D {\bf 66}, 123510 (2002).
   
\bibitem{Baumann:2015xxa} 
  D.~Baumann, H.~Lee and G.~L.~Pimentel,
  arXiv:1507.07250 [hep-th].
  
\bibitem{Benson:2014qhw} 
  B.~A.~Benson {\it et al.} [SPT-3G Collaboration],
  Proc.\ SPIE Int.\ Soc.\ Opt.\ Eng.\  {\bf 9153}, 91531P (2014).
  
\bibitem{footnote7}
  On the other hand, we note that any excess in radiation energy density could be due either to primordial gravitational waves or to additional light particles, not predicted by the Standard Model of particle physics.
  
\bibitem{footnoteENRICO}
  In the slow-roll approximation, one can write the fractional change of $\epsilon_H\equiv\epsilon$ per e-fold of inflation as
\begin{equation*}
\epsilon\approx\epsilon_\star + \frac{\dif\epsilon}{\dif N}\bigg|_\star\Delta N\,\,.
\end{equation*}
Recalling that
\begin{equation*}
-(n_\tu{s} - 1)|_\star = 2\epsilon_\star - \frac{1}{\epsilon_\star}\frac{\dif\epsilon}{\dif N}\bigg|_\star\,\,,
\end{equation*}
one finds $\Delta\epsilon/\epsilon_\star\approx(0.04 + r_\star/8)\times\Delta N\approx 0.04\times\Delta N$. For $\Delta N\approx\log k_\mathrm{LIGO}/k_\mathrm{CMB}\approx\log\num{d19}$, one has a fractional change in the slow-roll parameter $\epsilon$ of order $1.7$.
  
  

\bibitem{Remazeilles:2015hpa} 
  M.~Remazeilles, C.~Dickinson, H.~K.~K.~Eriksen and I.~K.~Wehus,
  arXiv:1509.04714 [astro-ph.CO].
  
\bibitem{Lasky:2015lej} 
  P.~D.~Lasky {\it et al.},
  arXiv:1511.05994 [astro-ph.CO].

\bibitem{Adshead:2010mc} 
  P.~Adshead, R.~Easther, J.~Pritchard and A.~Loeb,
  JCAP {\bf 1102}, 021 (2011).

\bibitem{Perotto:2006rj} 
  L.~Perotto, J.~Lesgourgues, S.~Hannestad, H.~Tu and Y.~Y.~Y.~Wong,
  JCAP {\bf 0610}, 013 (2006).
  
\bibitem{Bond:1998qg}
  J.~R.~Bond, A.~H.~Jaffe and L.~E.~Knox,
  Astrophys.\ J.\  {\bf 533} (2000) 19.
  
\bibitem{Bond:1997wr} 
  J.~R.~Bond, G.~Efstathiou and M.~Tegmark,
  Mon.\ Not.\ Roy.\ Astron.\ Soc.\  {\bf 291}, L33 (1997).
  
\bibitem{Efstathiou:1998xx}
  G.~Efstathiou and J.~R.~Bond,
  Mon.\ Not.\ Roy.\ Astron.\ Soc.\  {\bf 304} (1999) 75.
  
\bibitem{Verde:2005ff} 
  L.~Verde, H.~Peiris and R.~Jimenez,
  JCAP {\bf 0601}, 019 (2006).
  
\bibitem{Hobson:1998td}
  M.~P.~Hobson, A.~W.~Jones, A.~N.~Lasenby and F.~R.~Bouchet,
  Mon.\ Not.\ Roy.\ Astron.\ Soc.\  {\bf 300} (1998) 1.
  
\bibitem{Baccigalupi:2002bh}
  C.~Baccigalupi, F.~Perrotta, G.~De Zotti, G.~F.~Smoot, C.~Burigana, D.~Maino, L.~Bedini and E.~Salerno,
  Mon.\ Not.\ Roy.\ Astron.\ Soc.\  {\bf 354} (2004) 55.
  
\bibitem{Wandelt:2000av}
  B.~D.~Wandelt, E.~Hivon and K.~M.~Gorski,
  Phys.\ Rev.\ D {\bf 64} (2001) 083003.
  
\bibitem{Barnaby:2010vf} 
  N.~Barnaby and M.~Peloso,
  Phys.\ Rev.\ Lett.\  {\bf 106}, 181301 (2011).
  
\bibitem{Sorbo:2011rz} 
  L.~Sorbo,
  JCAP {\bf 1106}, 003 (2011).
  
\bibitem{Barnaby:2011vw} 
  N.~Barnaby, R.~Namba and M.~Peloso,
  JCAP {\bf 1104}, 009 (2011).
  
\bibitem{footnote-app3-1}
  We note that, since $f^L \neq f^R$, the primordial gravitational waves produced by this mechanism will be parity violating.

\end{thebibliography}
\end{document}